\documentclass[11pt,a4paper]{article}

\usepackage[utf8]{inputenc}
\usepackage[T1]{fontenc}
\usepackage{lmodern}
\usepackage[a4paper,margin=2.5cm]{geometry}
\usepackage{microtype}
\usepackage[english]{babel}
\usepackage{csquotes}

\usepackage[table]{xcolor}
\usepackage{amsmath}
\usepackage{amssymb}
\usepackage{graphicx}
\usepackage{booktabs}
\usepackage{array}
\usepackage{tabularx}
\usepackage{ragged2e}
\usepackage{multirow}
\usepackage{enumitem}
\usepackage{tikz}
\usepackage{xurl}
\usepackage[hidelinks]{hyperref}
\usepackage{stfloats}
\usepackage{placeins}
\usepackage{float}

\usepackage{titlesec}

\usepackage{longtable}

\definecolor{aiBlueDark}{HTML}{183A66}
\definecolor{aiBlueFill}{HTML}{EEF5FC}
\definecolor{aiGreenDark}{HTML}{2F6B3C}
\definecolor{aiGreenFill}{HTML}{EEF7EE}
\definecolor{aiPurpleDark}{HTML}{5D3F8C}
\definecolor{aiPurpleFill}{HTML}{F3EFFA}
\definecolor{aiOrangeDark}{HTML}{B45A1C}
\definecolor{aiOrangeFill}{HTML}{FFF4EA}
\definecolor{aiGrayDark}{HTML}{4A5568}

\usepackage{float}
\usepackage{xcolor}
\usepackage{tabularx}
\usepackage{booktabs}
\usepackage{colortbl}

\usepackage[font=small,labelfont=bf]{caption}

\titlespacing*{\section}{0pt}{16pt plus 2pt minus 1pt}{16pt plus 2pt minus 1pt}
\titlespacing*{\subsection}{0pt}{12pt plus 2pt minus 1pt}{12pt plus 2pt minus 1pt}
\titlespacing*{\subsubsection}{0pt}{10pt plus 1pt minus 1pt}{10pt plus 1pt minus 1pt}


\newcolumntype{L}[1]{>{\RaggedRight\arraybackslash}p{#1}}
\newcolumntype{Y}{>{\RaggedRight\arraybackslash}X}

\usetikzlibrary{positioning, arrows.meta, shapes.geometric, calc,
                fit, backgrounds, decorations.pathmorphing}

\definecolor{stageDark}{HTML}{0E1A2B}
\definecolor{stageBlue}{HTML}{2A6EBB}
\definecolor{stageTeal}{HTML}{1FA89C}
\definecolor{stageGreen}{HTML}{6FB23A}
\definecolor{stagePurple}{HTML}{7B4FA0}
\definecolor{stageOrange}{HTML}{E08A2C}
\definecolor{boxPre}{HTML}{E6EEFA}
\definecolor{boxCtrl}{HTML}{E2F2EE}
\definecolor{boxDec}{HTML}{F8ECDC}
\definecolor{linkGray}{HTML}{C7CCD3}

\definecolor{stageBlueDark}{HTML}{184A8C}
\definecolor{stageBlueFill}{HTML}{EAF2FB}
\definecolor{stageTealDark}{HTML}{0F8C95}
\definecolor{stageTealFill}{HTML}{E7F7F8}
\definecolor{stageGreenDark}{HTML}{2E7D32}
\definecolor{stageGreenFill}{HTML}{EEF7EE}
\definecolor{stagePurpleDark}{HTML}{6A3FB5}
\definecolor{stagePurpleFill}{HTML}{F3ECFA}

\title{AI-driven Software Development:\\
A Pragmatic Path to Agentic Development Processes}

\author{
Peter Mandl\\
Munich University of Applied Sciences\\
Munich, Germany\\
\texttt{peter.mandl@hm.edu}
\and
Paul Mandl\\
Findustrial GmbH\\
Sch{\"o}rfling am Attersee, Austria\\
\texttt{paul.mandl@findustrial.io}
}

\date{}

\begin{document}

\pagestyle{plain}
\maketitle
\thispagestyle{plain}

\begin{abstract}
Generative AI is increasingly transforming software development from
localized tool support into a form of development work that is more
deeply embedded in processes, tools, and organizational structures.
Its use is no longer limited to code completion, but affects
requirements, architecture, implementation, testing, review,
operations, and maintenance. Existing research presents a differentiated
picture. Productivity gains are possible, but depend strongly on
factors such as task type, codebase characteristics, and developers'
level of experience. At the same time, AI-generated artifacts require
additional control and governance mechanisms.

Building on these observations, this paper develops a pragmatic
organizing framework for the transition toward AI-driven Software
Development. The framework describes the progression from informal
and assistive AI use through integrated AI workflows toward controlled
agentic development processes. The focus is not on evaluating
individual tools or models, but on how AI can be embedded technically,
organizationally, and through quality assurance across central
software engineering activities. Particular importance is assigned to
the development of a harness that connects project context, tool
access, verification mechanisms, permissions, logging, and human
approval.

The paper draws on current research literature, practice-oriented
sources, established software engineering practices, and project
experience. To substantiate the framework, a mid-sized software
company is used as an exploratory case study. The case study serves to
assess the plausibility of the proposed organizing framework and
shows how technical prerequisites, governance requirements, design
practices, and transformation paths can be shaped in a concrete
organizational context. The paper thereby provides a conceptual basis
for further scholarly discussion and empirical investigation of
AI-driven Software Development.
\end{abstract}

\vspace{0.5em}
\vspace{1em}

\section{Introduction}

Advances in generative AI are reshaping how software is created, tested, documented, and
operated. In software-intensive organizations, this raises questions not only about suitable
tools but also about how they can be embedded within development processes. This paper
examines these developments under the term \emph{AI-driven Software Development} as an
overarching perspective on the use of artificial intelligence across the software development
lifecycle (SDLC).

The practical relevance is already visible in numerous industrial applications. Google
reported in 2024 that more than a quarter of its new code is generated by AI and subsequently
reviewed by engineers~\cite{google2024q3}. Amazon documents the migration of approximately
30,000 applications to Java~17 with measurable effort and cost reductions~\cite{amazon2024qdeveloper}.
Shopify has established AI use as a daily work expectation, linking it to resource requests
and reviews~\cite{shopify2025memo,theverge2025shopify}. A joint study by GitHub and Accenture
reports high Copilot adoption in professional development teams~\cite{github2024accenture}.

While these examples illustrate growing adoption, they are insufficient for a rigorous
assessment of effects and risks in real development organizations. Controlled experiments
and field studies report substantial time savings in some settings~\cite{peng2023copilot,cui2025highskilled}.
A randomized study with experienced open-source developers working in familiar repositories,
however, found longer task completion times under AI assistance~\cite{metr2025productivity}.
Security studies further indicate that AI-generated code can contain vulnerabilities and
should not be integrated into production systems without additional review
steps~\cite{pearce2022asleep,spracklen2024packagehallucinations}.

The central challenge therefore lies less in the isolated introduction of individual AI
tools than in their controlled integration. Much of the existing literature focuses on
individual activities such as code generation, requirements engineering, test support, or
automated bug fixing. Organizations additionally need guidance on how AI support can be
embedded technically, organizationally, and in regulatory terms to enable the controlled
adoption and evaluation of AI-driven Software Development.

This paper develops a pragmatic framework for the transition from selective AI support
to more integrated forms of AI-driven Software Development. The focus is not on evaluating
individual tools or models, but on how AI usage can be systematically organized and governed
across key software engineering activities---from assistive individual functions through
integrated workflows to more fully AI-shaped development and operational processes.

The paper is structured as follows. Section~\ref{sec:approach} describes the research
approach. Section~\ref{sec:related} positions the work within the research landscape.
Section~\ref{sec:ai_support_swe_activities} examines AI support across key software
engineering activities. Section~\ref{sec:transformation} presents the transformation model.
Section~\ref{sec:gestaltung} derives concrete design practices. Section~\ref{sec:caseA}
discusses the case study. Conclusions, limitations, and outlook follow.

\section{Research Approach}
\label{sec:approach}

\begin{figure}[t]
\centering
\begin{tikzpicture}[
    font=\small,
    >=Stealth,
    stepbox/.style={
        rounded corners=6pt,
        line width=0.7pt,
        inner sep=5pt,
        align=center,
        minimum width=11.4cm,
        text width=10.6cm,
        minimum height=1.35cm,
        text=black
    },
    stepnum/.style={
        circle,
        fill=white,
        line width=0.6pt,
        minimum size=6mm,
        inner sep=0pt,
        font=\scriptsize\bfseries
    },
    arr/.style={
        -{Stealth[length=2.0mm]},
        line width=0.7pt,
        draw=aiBlueDark,
        shorten >=1pt,
        shorten <=1pt
    },
    note/.style={
        font=\footnotesize,
        text=aiGrayDark
    }
]

\node[
    stepbox,
    draw=aiGreenDark,
    fill=aiGreenFill
] (base) at (0,0) {%
    \textbf{\textcolor{aiGreenDark}{Establish Foundations}}\\[2pt]
    State of Research \;|\; Project Experience
};
\node[stepnum, draw=aiGreenDark, text=aiGreenDark]
    at ([xshift=0.42cm,yshift=-0.42cm]base.north west) {1};

\node[
    stepbox,
    draw=aiBlueDark,
    fill=aiBlueFill
] (act) at (0,-2.15) {%
    \textbf{\textcolor{aiBlueDark}{Analyze AI Support in Software Engineering}}\\[2pt]
    Requirements Engineering \;|\; Implementation \;|\; Test/Review \;|\; Deployment/Operations \;|\; Maintenance
};
\node[stepnum, draw=aiBlueDark, text=aiBlueDark]
    at ([xshift=0.42cm,yshift=-0.42cm]act.north west) {2};

\node[
    stepbox,
    draw=aiPurpleDark,
    fill=aiPurpleFill
] (trans) at (0,-4.30) {%
    \textbf{\textcolor{aiPurpleDark}{Derive Transformation Model}}\\[2pt]
    AI-assisted $\rightarrow$ AI-integrated $\rightarrow$ AI-driven\\[1pt]
    Tool Support \;|\; Governance \;|\; Productivity Measurement
};
\node[stepnum, draw=aiPurpleDark, text=aiPurpleDark]
    at ([xshift=0.42cm,yshift=-0.42cm]trans.north west) {3};

\node[
    stepbox,
    draw=aiOrangeDark,
    fill=aiOrangeFill
] (cases) at (0,-6.45) {%
    \textbf{\textcolor{aiOrangeDark}{Integrate Exploratory Case Study}}\\[2pt]
    Mid-Sized Software Company
};
\node[stepnum, draw=aiOrangeDark, text=aiOrangeDark]
    at ([xshift=0.42cm,yshift=-0.42cm]cases.north west) {4};

\draw[arr] (base.south) -- (act.north);
\draw[arr] (act.south) -- (trans.north);
\draw[arr] (trans.south) -- (cases.north);

\end{tikzpicture}
\caption{Research approach of this paper.}
\label{fig:research_approach}
\end{figure}
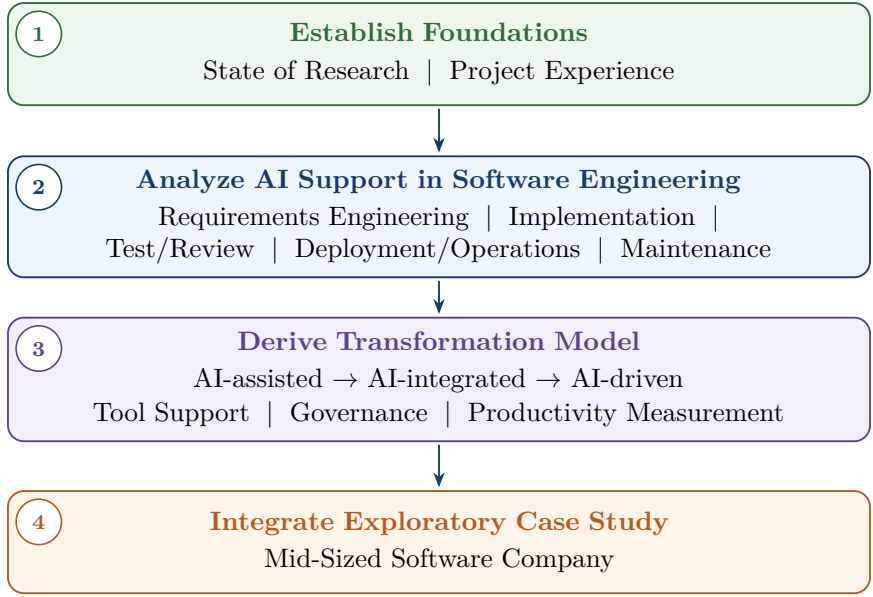

The paper follows a multi-stage, conceptual approach that combines literature analysis,
project experience, model development, and an exploratory case study.
Figure~\ref{fig:research_approach} summarizes the four steps.

\emph{Establish foundations.}
The foundational layer combines scientific studies on AI-assisted software development with
project experience from the authors' professional context. The literature review was conducted
exploratively and iteratively across ACM Digital Library, IEEE Xplore, arXiv, and Google
Scholar, supplemented by practitioner publications from cloud, platform, and tool providers.
Search terms included \emph{LLMs for Software Engineering}, \emph{AI Coding Assistants},
\emph{AI-assisted Software Engineering}, \emph{AI-driven Software Development},
\emph{AI-native Software Engineering}, \emph{Human-AI Collaboration},
\emph{Agentic Software Engineering}, \emph{Program Repair}, \emph{DevOps Metrics},
and \emph{AI Governance}. Given that terminology in this field is not yet consolidated, these
terms were used as sensitizing concepts rather than disjoint analytical categories.

\emph{Analyze AI support in software engineering.}
The second step examines how generative AI can support key software engineering activities,
from requirements engineering and implementation through testing, review, deployment,
operations, and maintenance. The goal is not to evaluate individual tools but to identify
recurring patterns by synthesizing findings from research, practitioner reports, and project
experience.

\emph{Derive transformation model.}
On this basis, a pragmatic transformation model is developed that distinguishes between
AI-assisted, AI-integrated, and AI-driven Software Development, connecting these stages to
tool support, governance mechanisms, and productivity measurement. The model is intended as
a structuring framework rather than a normative maturity model.

\emph{Integrate exploratory case study.}
To ground the framework, an exploratory case study of a mid-sized custom software company
is presented. It is not intended to validate the model empirically but to contextualize the
proposed approach. The analysis draws on publicly available information, observable platform
structures, typical development and operations patterns, and the authors' project experience.

\definecolor{boxBorder}{gray}{0.25}
\definecolor{boxFillA}{gray}{0.96}
\definecolor{boxFillB}{gray}{0.94}
\definecolor{boxFillC}{gray}{0.92}
\definecolor{boxFillD}{gray}{0.95}

\section{Background and Related Work}
\label{sec:related}

This section situates the state of research and technology relevant to AI-driven Software Development. It first distinguishes central terms and related concepts. It then examines large language models and specialized code models as the technical foundation of current AI support in software engineering. In this context, it also discusses the required harness and tool integration. In this paper, the term \emph{harness} does not refer to a specific vendor or tool, but to the technical, organizational, and procedural embedding of AI systems into project context, development tools, execution environments, verification mechanisms, permission structures, and human approval processes. Building on this, the section situates related transformation and lifecycle approaches and summarizes empirical findings on human-AI collaboration, productivity, quality, security, and agentic systems. The aim is to identify the key reference points for the organizing framework developed in this paper.

\subsection{Conceptual Positioning}

Current literature on AI-assisted software development uses several overlapping terms.
\emph{LLM4SE} (\emph{Large Language Models for Software Engineering}) denotes the research
field concerned with applying large language models to software engineering
tasks~\cite{hou2024llmse,fan2023survey}. \emph{Generative software development} refers in
classical software engineering to model- and domain-specific language-based artifact
generation~\cite{rumpe2010generative}; in the context of generative AI it additionally
covers code, tests, specifications, and documentation produced by large language
models~\cite{jiang2024codegeneration,zhang2024agilegen}. \emph{AI-assisted Software
Engineering} describes a primarily assistive usage in which AI supports individual tasks,
whereas \emph{AI-native} or \emph{AI-first Software Engineering} implies deeper process and
organizational changes~\cite{hassan2024ainative,li2025aiteammates}. \emph{Agentic Software
Engineering} highlights systems that plan tasks, use development tools, and iteratively
verify results~\cite{yang2024sweagent,hong2024metagpt,wu2024autogen,wang2024openhands}.

In this paper, \emph{AI-driven Software Development} serves as the overarching working
term, encompassing software development practices in which AI substantially supports,
extends, or partly automates activities across the SDLC. The proposed transformation
framework distinguishes three usage forms: \emph{AI-assisted}, \emph{AI-integrated}, and
\emph{AI-driven} development. The related terms listed above are understood not as competing
categories but as adjacent perspectives on different expressions of AI-driven Software
Development.

\subsection{LLMs and Code Models in Software Engineering}

The technical foundation of most current approaches consists of large language models and
code-specialized models. Beyond proprietary frontier systems, openly available and specialized
code models are gaining importance. Code~Llama~\cite{roziere2023codellama} and
StarCoder~\cite{li2023starcoder} represent key reference points.

In practice, different model classes are employed. Proprietary frontier models such as
GPT-4.1 and GPT-5~\cite{openai2025gpt41,openai2025gpt5}, Claude Sonnet and Claude
Opus~\cite{anthropic2025claude4}, and Gemini~2.5 Pro~\cite{google2025gemini25pro} are
routinely used for software development tasks. Specialized code models such as
Codestral~\cite{mistral2024codestral}, DeepSeek-Coder~\cite{guo2024deepseekcoder}, and
Qwen2.5-Coder~\cite{hui2024qwen25coder} are designed specifically for code generation,
completion, understanding, reasoning, and repair.

For organizations with high data privacy and auditability requirements, the distinction
between proprietary, openly available, and locally deployable models is at least as
important as raw model performance, since it directly affects governance, operational risk,
and compliance requirements.

\subsection{Harness, Tool Integration, and Development Environments}
\label{sec:harness}

The impact of large language models and code models in software
engineering does not depend on model performance alone, but also
substantially on their embedding in development environments and
processes. The previously introduced term \emph{harness} refers to
the technical, organizational, and procedural environment through
which a model is connected to project context, tools, and control
mechanisms. This includes, in particular, access to repositories,
issues, documentation, build and test systems, as well as mechanisms
for logging, permission boundaries, quality assurance, and human
approval~\cite{hashimoto2026aiadoption,openai2026harness,bockeler2026harness,zhong2026aiharness}.

A harness translates model capabilities into usable development
workflows. It determines which information is made available to a
model, which actions it is allowed to perform, how results are
verified, and at which points human decisions remain necessary. It
therefore forms an intermediate layer between isolated model use and
productive integration into software development processes. For
AI-driven Software Development, this layer is central because it is
only through such embedding that generated suggestions become
traceable, verifiable, and controllable contributions to the
development process.

This harness perspective is increasingly reflected in current
development tools and agent-oriented coding environments. Tools such
as Claude Code, Amazon Q Developer, Cursor, GitHub Copilot, and
cloud-based coding agents connect language models with repository
context, file editing, command execution, test execution,
documentation, and review workflows
~\cite{anthropicclaudecode,amazon2024qdeveloper,cursoragent,githubcopilotdocs,githubcloudagent}.
They therefore illustrate that the practical value of AI support does
not arise from model access alone, but from the surrounding
environment that structures context provision, tool use, quality
checks, permissions, and human oversight. At the same time, these
tools underline that harness design should be treated as an explicit
engineering and governance task rather than as a purely technical
integration detail.

\subsection{Related Approaches to AI-Assisted Software Development}

The organizing framework proposed in this paper is situated within
a broader landscape of approaches to the AI-assisted further
development of software engineering. These approaches, however, do
not pursue the same objective. Some primarily describe how the
software lifecycle changes through generative AI, while others focus
more strongly on organizational prerequisites, role models, or tool
pipelines.

A phase-oriented perspective can be found, for example, in the
AI-driven Development Life Cycle proposed by AWS. In this view, AI is
understood as part of the entire development process, ranging from
the early clarification of business requirements through construction
and validation to operations~\cite{aws2025addlc}. The DORA AI
Capabilities Model by Google takes a different perspective. Rather
than proposing a process model, it asks which capabilities
organizations need for the effective use of AI-assisted Software
Development. These include, among others, clear guidelines, reliable
access to data and context, platform maturity, and robust
engineering practices~\cite{dora2025aicapabilities}.

Research on individual lifecycle activities further illustrates that
AI support extends beyond implementation. In requirements engineering,
Zadenoori et al.\ show that large language models can support
elicitation, validation, and specification analysis
~\cite{zadenoori2025requirements}. This strengthens the view that
AI-assisted software development must be considered across the
lifecycle rather than as a code-generation phenomenon alone.

A more future-oriented perspective is articulated in work on
\emph{AI-native Software Engineering} and \emph{Software Engineering
3.0} (SE~3.0). These concepts refer to a vision in which software
development is driven more strongly by business and technical intent,
and in which AI systems are integrated as collaborative team partners
into development environments, toolchains, and runtime processes
~\cite{hassan2024ainative,li2025aiteammates}.

Within these discussions, the term \emph{Zero-Touch Engineering}
appears as a boundary case of extensive agentic automation, in which
AI systems independently prepare or execute large portions of
analysis, implementation, testing, deployment, and operations. It is
adjacent to work on Zero-Touch DevOps and agentic software
engineering systems
~\cite{sethupathy2023zerotouchdevops,zhang2024agents,wang2024openhands,zhong2026aiharness}.
For this paper, Zero-Touch Engineering is relevant as a boundary
concept indicating where agentic development environments could
ultimately lead. It is not treated as a standalone transformation
stage, since its practical realization continues to depend on human
goal-setting, verification, approval, and organizational control.

Analytical and practitioner-oriented work, for example on TuringBots,
AI-augmented software engineering, or AI-assisted software delivery,
further emphasizes changes in tools, roles, and platforms. The focus
of these contributions is less on a fully developed scientific
transformation model than on the observation that AI support is
increasingly connecting multiple activities across the software
lifecycle. As a result, developer roles shift more strongly toward
orchestration, control, and system design
~\cite{forrester2024turingbots,gartner2025softwaretrends,mckinsey2025softwarevalue,thoughtworks2024aitools}.

What these approaches have in common is that generative AI is not
understood as an isolated individual tool. Its value emerges only
through its embedding in the development process. The present paper
adopts this basic assumption, but focuses more specifically on the
logic of transformation. It describes the transition from AI-assisted
through AI-integrated to AI-driven Software Development as a
pragmatic organizing framework for organizations that seek to
classify and further develop their use of AI more systematically.

\subsection{Human-AI Collaboration}

Several studies show that the impact of generative AI in software
engineering does not depend on model quality alone. Based on a
workshop with software engineers, Hamza et al.\ describe a shift in
the role of AI from a mere tool toward a collaborative partner
~\cite{hamza2023humanai}. Human oversight remains indispensable,
especially for complex or safety-critical tasks.

Treude et al.\ develop a taxonomy of human-AI interaction in software
engineering, covering a broad spectrum of use cases from querying
existing codebases to debugging, planning, review, and documentation
tasks~\cite{treude2025taxonomy}. Barke et al.\ examine the practical
use of GitHub Copilot and identify both acceleration-oriented use in
familiar tasks and exploratory use when developers work with
unfamiliar technologies or interfaces~\cite{barke2023groundedcopilot}.

Against this background, Zero-Touch Engineering should be understood
as a boundary case rather than as a realistic normal form of software
development. Work on Zero-Touch DevOps and agentic software
engineering systems shows that larger parts of the development
process can be automatically prepared or executed
~\cite{sethupathy2023zerotouchdevops,zhang2024agents,wang2024openhands}.
However, empirical findings on human-AI collaboration suggest that
human goal setting, domain-specific assessment, security review, and
approval remain central components of productive software
development. Zero-touch therefore describes less the disappearance of
human responsibility than a shift of human work toward steering,
review, and control.

Vaithilingam et al.\ analyze early experiences with AI-based tools
for code generation~\cite{vaithilingam2022expectation}. Mozannar et
al.\ examine the additional cognitive effort involved in reading and
reviewing AI-generated suggestions~\cite{mozannar2024reading}. Russo
addresses organizational factors influencing the adoption of
generative AI in development organizations~\cite{russo2024adoption}.
Khemka and Houck complement this perspective with developers'
expectations and concerns, particularly regarding trust,
transparency, and potential loss of control~\cite{khemka2024support}.

\subsection{Productivity and Empirical Impact}

The empirical evidence on productivity effects is mixed. Peng et al.
report a substantial reduction in task completion time for programming
tasks in a controlled experiment with GitHub Copilot~\cite{peng2023copilot}.
Cui et al.\ also observe productivity gains in field studies, but
emphasize the strong context dependence of the findings
~\cite{cui2025highskilled}.

An important counterpoint is provided by the METR study. In this
study, experienced open-source developers worked on repositories
familiar to them and, with AI support, required more time on average
than without it~\cite{metr2025productivity}. The authors attribute
this, among other factors, to additional review effort and the need to
correct generated suggestions.

Liang et al.\ examine the usability of AI-based programming assistants
in a large-scale developer survey~\cite{liang2024usability}.
Sergeyuk et al.\ analyze real usage patterns in development
environments~\cite{sergeyuk2024practice}. Industry reports by GitHub
and Microsoft on code quality, as well as the annual DORA State of
DevOps Report, complement these findings from an application-oriented
perspective~\cite{github2024codequality,dora2024stateofdevops}.

Overall, the available studies suggest that productivity gains depend
on several factors, including task complexity, the experience level of
the users, the codebase, process integration, and review effort. The
amount of generated code is not a meaningful success measure. More
suitable indicators include lead time, defect density, maintainability,
security, and developer satisfaction.

\subsection{Security, Correctness, and Supply Chain Risks}

Several studies document security and quality problems in AI-generated code. Pearce et al.\
examined GitHub Copilot in security-critical scenarios and found a significant share of
vulnerable suggestions~\cite{pearce2022asleep}. Sandoval et al.\ analyzed the security
consequences of AI-based coding assistants in a user study~\cite{sandoval2023lostatc}, and
Khoury et al.\ report similar findings for ChatGPT-generated code~\cite{khoury2023securechatgpt}.

Assessing functional correctness remains challenging. Liu et al.\ show with EvalPlus that
standard benchmarks such as HumanEval~\cite{chen2021codex} overestimate actual model
performance when stricter test cases are applied~\cite{liu2024evalplus}. Ouyang et al.\
highlight the non-determinism of code generation as an additional complication for
reproducibility~\cite{ouyang2024nondeterminism}. Spracklen et al.\ document hallucinated
package names and the resulting supply chain risks~\cite{spracklen2024packagehallucinations},
and Zhang et al.\ identify systematic differences between model providers in code
generation~\cite{zhang2025providerbias}.

\subsection{Project-Level Evaluation and Agentic Systems}

A key challenge for AI-driven Software Development is that many
standard benchmarks only approximate real software engineering work.
They often focus on isolated programming tasks, while productive
development requires understanding repository context, modifying
several files, running tests, interpreting failures, and integrating
changes into existing architectures and workflows. Project-level
evaluation therefore becomes essential for assessing whether AI
systems can contribute reliably to realistic development tasks.

SWE-bench addresses this issue by evaluating language models on
real-world software engineering problems derived from GitHub issues
and pull requests~\cite{jimenez2024swebench}. This shifts the focus
from isolated code completion toward repository-level tasks. Related
agentic approaches, such as SWE-agent, show how language models can
interact with files, tests, and development tools in order to solve
software engineering tasks more autonomously~\cite{yang2024sweagent}.
Other systems, including MetaGPT, AutoGen, and OpenHands, explore
multi-agent coordination, conversational agent frameworks, and open
platforms for software development agents
~\cite{hong2024metagpt,wu2024autogen,wang2024openhands}. Broader
surveys on autonomous LLM agents position these systems as an
emerging field that extends classical coding assistance toward
partially autonomous development workflows~\cite{zhang2024agents}.

Beyond benchmark-oriented code generation, recent work also addresses
evaluation and repair tasks more directly. Jiang et al.\ survey code
generation with language models, while Chen et al.\ compare
evaluation methods for AI-based code generation
~\cite{jiang2024codegeneration,chen2024codeevaluation}. Xia et al.
show that pre-trained models can match or surpass established
automated program repair approaches in several scenarios
~\cite{xia2023apr}. These findings support the shift from isolated
code suggestions toward project-related development tasks that
require testing, repair, and evaluation.

For the transformation perspective developed in this paper, these
results are important for two reasons. First, they show that agentic
systems can extend AI support beyond suggestion generation toward
task-oriented interaction with development environments. Second, they
make clear that such systems require stronger safeguards than
traditional coding assistants. Traceability, reproducibility,
permission boundaries, test execution, logging, and human approval
are not secondary implementation details, but prerequisites for using
agentic systems in productive software engineering contexts.

\subsection{Summary}

The overall picture confirms that AI-driven Software Development
cannot be reduced to code generation. Large language models and
specialized code models increasingly support activities across the
software lifecycle, including requirements engineering, architecture,
implementation, testing, review, operations, and maintenance. At the
same time, the benefits and risks of AI use depend strongly on task
type, codebase characteristics, process integration, developer
experience, and review effort.

The discussion is therefore shifting from isolated assistance toward
the integration of AI into development environments, quality
assurance, and governance structures. Harness design, tool
integration, human oversight, verification, and security review
emerge as central prerequisites for productive use. Agentic systems
extend this development by introducing new forms of partial
automation, but they also raise additional requirements for
traceability, accountability, auditability, and organizational
control.

Appendix Tables~\ref{tab:terminology_ai_driven} and
\ref{tab:related_evidence_risks} summarize key terms and selected
research and practice sources. On this basis, the following section
examines how AI support can be classified along key software
engineering activities before being transferred into a pragmatic
transformation model.

\section{AI Support for Key Software Engineering Activities}
\label{sec:ai_support_swe_activities}

Following the conceptual and methodological positioning in the previous sections, this section turns to the concrete activities of software engineering. Generative AI affects not just a single stage of development, such as code generation, but can support different phases of the software lifecycle. It should therefore not be understood merely as a tool for individual tasks, but as part of a broader development and learning process.

\subsection{Requirements Analysis and Specification}

In requirements analysis and specification, the strengths of large language models are most visible in structuring, condensing, and generating alternatives. Language models can transform free-text input from interviews, tickets, or workshops into user stories and acceptance criteria, make implicit assumptions explicit, and suggest alternative formulations. A systematic literature review by Zadenoori et al.\ confirms these application patterns in requirements engineering~\cite{zadenoori2025requirements}; lifecycle-wide surveys also classify requirements-related tasks as a growing field of application~\cite{hou2024llmse,fan2023survey}. However, the domain-specific validation and legitimation of requirements remain the responsibility of product owners, business stakeholders, and other accountable domain roles.

Quality assurance of AI-generated requirements artifacts is particularly important. Requirements generated or reformulated with large language models may appear plausible while still being incomplete, ambiguous, difficult to verify, or insufficiently aligned with existing constraints. Recent work on LLMs in requirements engineering indicates that the use of such models continues to raise challenges related to reproducibility, evaluation, domain adaptation, and integration into robust workflows~\cite{norheim2024llmre,zadenoori2025requirements}. For practical use, AI-generated requirements should therefore not be treated as finished specifications, but as drafts that require systematic review and domain-specific approval.

One approach to quality assurance is to structure requirements through templates, controlled language, and explicit acceptance criteria. The \emph{Easy Approach to Requirements Syntax} (EARS) shows that natural-language requirements can be formulated more precisely, consistently, and verifiably through simple sentence patterns~\cite{mavin2009ears}. For AI-assisted requirements work, this means that models should not merely generate free-text suggestions, but should make conditions, triggers, system responses, exceptions, and measurable acceptance criteria explicit. Requirements and test cases can be used reciprocally for quality assurance: Acceptance and system tests can be derived from requirements, while existing or newly generated test cases can help reveal missing conditions, unclear expectations, and incomplete requirements~\cite{fischbach2022cira,wang2019acceptancetests}. Such a review framework can be part of a harness without replacing domain-specific evaluation and approval of the requirements.

\subsection{Architecture Design and Implementation}

In architecture design and implementation, AI supports not only the generation of individual code fragments but increasingly also the handling of coherent technical tasks. These include sketching alternative architecture and interface variants, deriving adapter logic, refactoring, bug fixing, test preparation, and changes across multiple files~\cite{peng2023copilot,cui2025highskilled,barke2023groundedcopilot}. Surveys on code generation and evaluation provide a methodological classification of these capabilities~\cite{jiang2024codegeneration,chen2024codeevaluation}. Particularly for recurring development tasks and when working with unfamiliar frameworks, substantial productivity gains can result.

For architecture design and implementation, the harness is primarily the environment in which AI-generated suggestions and changes are connected with repository context, development tools, build and test systems, and review processes. It determines which parts of the codebase an AI system can take into account, which architecture and implementation decisions it may prepare, which changes it can suggest or execute, and how these changes are reviewed before adoption.

Value is created only when generated designs and code changes are integrated into existing architecture, build, quality, and security requirements. Architecture decisions often affect long-term system properties such as maintainability, extensibility, performance, security, and technical debt. They therefore cannot be derived from locally plausible model suggestions alone, but require contextual evaluation against existing architectural constraints and organizational development goals. Zhang et al.\ also point to differences between model providers that should be considered when selecting tools~\cite{zhang2025providerbias}. At the same time, the focus of many developer tasks shifts from
pure syntax production toward problem decomposition, context
management, and verification. Mozannar et al.\ note that reading and
checking AI-generated suggestions can create additional cognitive
effort~\cite{mozannar2024reading}. In stable development processes
with clear quality and review mechanisms, this effort may be reduced
and, depending on the task context, partly offset by productivity
gains in other development phases.

\subsection{Test and Review}

AI can add the most value in testing and review where artifacts are currently produced
incompletely, late, or manually. Models can propose unit tests, test data, and pull-request
summaries, making coverage gaps visible earlier, preparing reviews more thoroughly, and
accelerating routine checks. For legacy code or poorly documented interfaces, AI can help
describe existing behavior and derive initial regression tests.

This value only materializes if AI-generated test artifacts are themselves reviewed. Multiple
studies document security and correctness problems in AI-generated
code~\cite{pearce2022asleep,sandoval2023lostatc,khoury2023securechatgpt,liu2024evalplus,spracklen2024packagehallucinations}.
Without execution, coverage assessment, regression testing, and human
review, AI-generated tests may provide limited meaningful coverage,
and reviews risk becoming superficial. The non-determinism of model
outputs creates additional requirements for reproducibility and
logging~\cite{ouyang2024nondeterminism}. The harness connects
verification mechanisms with AI-assisted artifact generation, but it
replaces neither quality gates nor human judgment.

\subsection{Deployment and Operations}

In deployment and operations, AI primarily supports automation, analysis, and operational
assistance. Language models are already used to configure deployment pipelines, explain
infrastructure definitions, generate release notes, and identify misconfigurations in build
and deployment processes. In CI/CD environments, AI can assist with analyzing failed builds,
prioritizing error messages, and preparing rollback or recovery
measures. During live operations, AI helps through condensation,
classification, and prioritization. It can summarize logs, classify
incidents, and draft runbooks, thereby enabling faster pattern
recognition and earlier incident containment. Interventions affecting
production should initially be limited to suggestions, with permission
scoping, audit logs, and approval processes remaining necessary to
secure production-proximate actions.

\subsection{Maintenance}

In maintenance, the value of generative AI lies primarily in analyzing existing systems and
supporting long-term upkeep. Models can explain legacy code, surface technical debt, estimate
the impact of changes, and assist with bug fixes and minor refactorings---particularly
valuable in historically grown systems with limited documentation.

A growing application area is AI-assisted migration and modernization of legacy code. Work
on mainframe and COBOL systems shows that large language models can be applied to code
comprehension, summarization, translation to modern programming languages, and refactoring
preparation~\cite{xmainframe2024,ibm2024coboljava,gandhi2024coboljava}. Case studies on
PL/SQL-to-Java migration demonstrate that LLMs can assist in transforming large legacy
systems when domain models, examples, target architectures, and validation mechanisms are
incorporated into the process~\cite{solovyeva2025plsqljava}. These studies consistently
show that legacy migration cannot be reduced to code translation: semantic equivalence,
testability, target architecture integration, and expert review remain decisive.

Xia et al.\ demonstrate that pre-trained models can match or exceed
established automated program repair approaches in several scenarios
~\cite{xia2023apr}. Changes should nonetheless remain small,
test-secured, and reversible, especially for refactorings,
migrations, and security-sensitive modifications. Generated
documentation must be versioned and regularly reconciled with the
actual system state~\cite{github2024codequality}.

The activity-level analysis shows that AI support generates different
potentials, risks, and control requirements depending on the lifecycle
phase. While individual tasks can already be supported in isolation,
comprehensive use across software development activities requires
additional technical, organizational, and governance prerequisites.
These prerequisites form the basis for the transformation model
presented next.

\section{Transformation Model for AI-driven Software Development}
\label{sec:transformation}

Building on the activity analysis, this section describes a possible transition from a well-organized, quality-assured development organization toward assistive, integrated, and agentic AI usage. The stages and transitions are presented as a pragmatic, practice-oriented framework intended to help organizations position their AI usage and to invite further discussion. Effort shifts and productivity measurement are addressed within this model; concrete design practices follow in Section~\ref{sec:gestaltung}.

\begin{figure}[t]
\centering
\includegraphics[width=\textwidth]{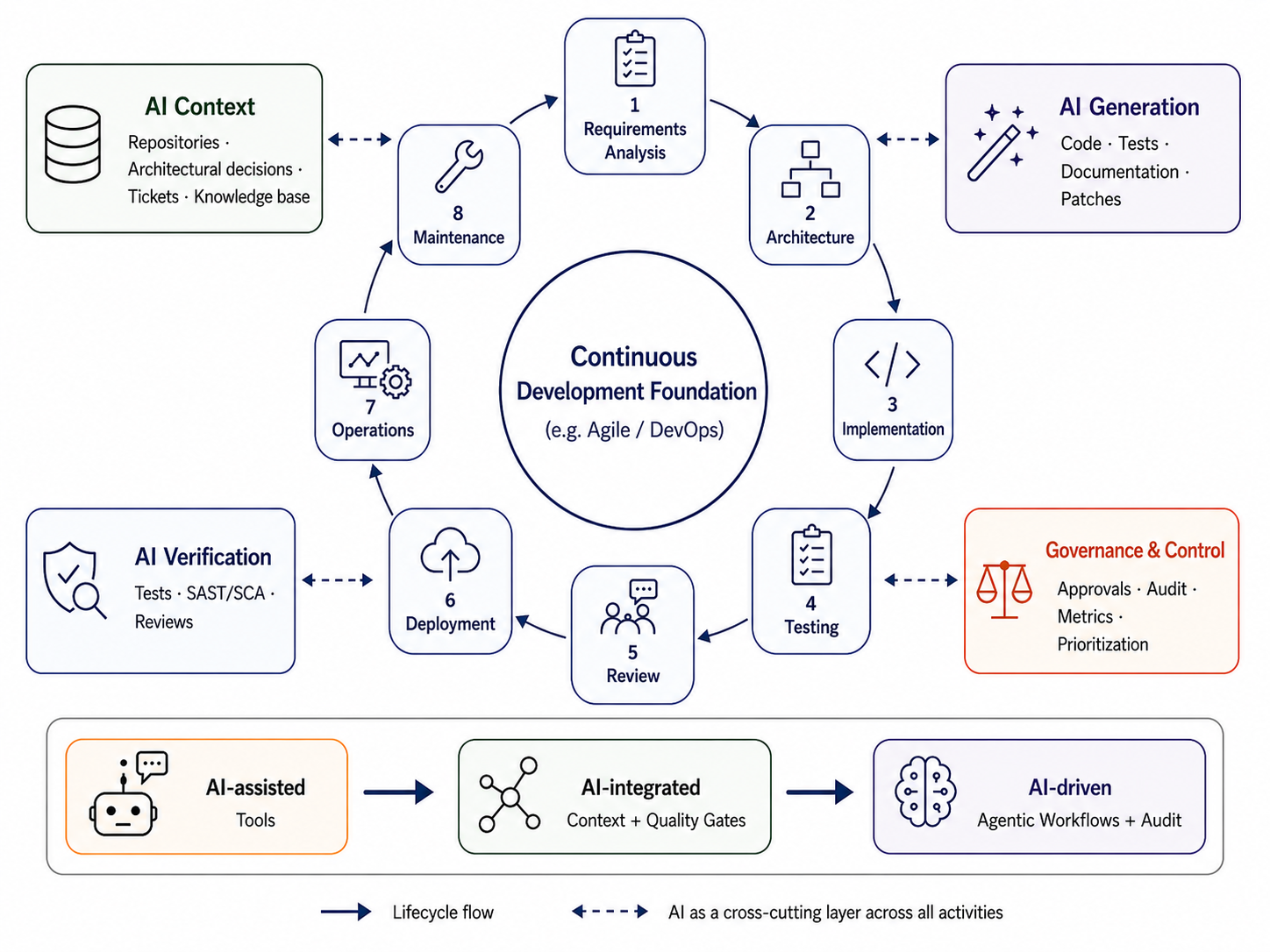}
\caption{Software engineering lifecycle and transformation logic in the context of AI-driven Software Development. The figure connects activities from requirements analysis through maintenance with the cross-cutting areas of AI Context, AI Generation, AI Verification, and Governance and Control. The lower bar illustrates the transformation stages from AI-assisted through AI-integrated to AI-driven Software Development.}
\label{fig:ai_lifecycle_transformation}
\end{figure}

Figure~\ref{fig:ai_lifecycle_transformation} presents the overarching
model. It shows the software engineering lifecycle as a cyclical
process from requirements analysis through architecture design,
implementation, testing, and deployment to operations and
maintenance. This process is centered on a consistent development
base that may be agile, DevOps-oriented, phase-based, or hybrid. What
matters is not the specific process model, but that requirements,
architecture design decisions, implementation, tests, releases,
operations, and maintenance are traceable and connected.

Four cross-cutting AI functions operate across all lifecycle
activities. \emph{AI Context} encompasses the project knowledge made
accessible to AI systems, including repositories, architecture design
decisions, tickets, and knowledge bases. \emph{AI Generation}
represents AI-assisted artifact production, including code, tests,
documentation, and patches. \emph{AI Verification} covers
project-level checking mechanisms, including tests, SAST/SCA, and
reviews. \emph{Governance and Control} describes the organizational
framework within which AI-generated artifacts are controlled and
integrated into development and operational processes. In this sense,
these areas can be understood as the harness through which AI systems
are embedded in development processes, supplied with context, oriented
toward verifiable artifacts, and organizationally controlled.

The lower bar of the figure further illustrates the transformation
logic used in this paper. It distinguishes the development from
AI-assisted through AI-integrated to AI-driven Software Development.
The following subsections explain the starting point, the proposed
transformation stages, the transition logic, associated effort
shifts, and the measurement of productivity.

\subsection{Starting Point: Informal and Heterogeneous Use}

In many organizations, AI adoption does not begin as a planned transformation program but as
informal, heterogeneous use of individual tools. Developers adopt Copilot, ChatGPT, Claude
Code, Cursor, or comparable systems without systematically adapting development processes,
quality assurance, data privacy handling, license review, or governance. This starting
point is not yet an integration stage; it describes a typical pre-phase in which AI is used
but organizationally only loosely framed.

Empirical studies show that usage patterns, tool selection, and expectations vary
considerably between teams and projects~\cite{liang2024usability,sergeyuk2024practice},
generating selective productivity gains alongside risks such as unreviewed code, privacy
concerns, unclear licensing, and uncertain accountability for AI-generated artifacts.
Russo shows that adoption and impact depend substantially on training, governance, and
platform maturity~\cite{russo2024adoption}, and Khemka and Houck identify trust, transparency,
and controllability as central adoption conditions from a developer perspective~\cite{khemka2024support}.
A structured transition model can help organizations treat AI adoption as a coordinated
change in practices, tools, and accountabilities.

\subsection{Transformation Stages}

The proposed framework does not describe strictly sequential maturity levels. In practice,
different usage forms often coexist: a team may use AI assistively for code completion while
simultaneously piloting agents for test generation or pull-request preparation. The relevant
question is therefore not formal stage assignment but how deeply AI is embedded in processes,
quality assurance, and accountability structures.

Table~\ref{tab:maturity} summarizes the three proposed stages. The presentation is a
pragmatic structuring proposal; it draws on the lifecycle, capability, and vision approaches
compared in Section~\ref{sec:related} and condenses them into an organization-level
transition logic.

The transitions between stages mark key prerequisites for the next integration level:
approved AI tools as the basis for governed assistance; context integration and quality gates
for integrated use; and agentic workflows with auditability for AI-driven development. With
the integrated stage, a harness gradually takes shape that connects context provisioning,
tool use, verification, and governance.
\begin{center}
\captionof{table}{Proposed Transformation Stages}
\label{tab:maturity}

\scriptsize
\renewcommand{\arraystretch}{1.25}
\setlength{\tabcolsep}{5pt}
\rowcolors{2}{gray!8}{white}

\begin{tabularx}{\columnwidth}{L{0.7cm} L{2.4cm} Y Y Y}
\toprule

\textbf{Stage} &
\textbf{Usage Mode} &
\textbf{Characteristics} &
\textbf{Practices} &
\textbf{Challenges}
\tabularnewline

\midrule

1 &
AI-assisted Development &
AI supports individual tasks: code, tests, documentation, explanations. &
Approved tools, usage guidelines, mandatory reviews, training. &
Heterogeneous adoption and unreviewed outputs.
\tabularnewline

2 &
AI-integrated Development &
AI is embedded in development sub-processes. &
Context systems, guidelines, test scenarios, quality gates. &
Governance, integration, and process alignment.
\tabularnewline

3 &
AI-driven or Agentic Development &
Agents support development, testing, operations, and maintenance. &
Audit logs, security mechanisms, role models, human approval. &
Control, traceability, and accountability.
\tabularnewline

\bottomrule
\end{tabularx}
\end{center}

Modern software development continues to rest on clear requirements, modular architecture,
automated testing, CI/CD, reviews, and clearly assigned ownership. These foundations are
especially mature in agile and DevOps-oriented organizations, but can be realized in
phase-based or hybrid processes as well. Where they are absent, AI can amplify existing
weaknesses rather than compensate for them.

\subsection{Transition Logic}

The proposed transition builds on established development practice and insights from
human-AI interaction and agentic software engineering. Similar premises appear in the AWS
AI-driven Development Life Cycle~\cite{aws2025addlc}, in work on the future of AI-assisted
software development~\cite{terragni2025future}, and in studies on human-AI
collaboration~\cite{treude2025taxonomy,hamza2023humanai}.

At the assistive stage, the goal is to introduce AI tools in a controlled manner within
existing processes: clear usage guidelines, training, data privacy review, and mandatory
reviews. Security studies on Copilot- and ChatGPT-generated code show that these safeguards
are needed from the outset~\cite{pearce2022asleep,sandoval2023lostatc,khoury2023securechatgpt}.

The integrated stage requires AI systems to work with project-specific
context. Knowledge from repositories, ticket systems, and project
documentation must be accessible and current.
Verification requirements also increase. Public benchmarks and
isolated test runs provide only limited evidence for assessing the
quality of generated solutions
~\cite{liu2024evalplus,ouyang2024nondeterminism}, making automated
tests, static analysis, and reusable evaluation cases necessary. This
combination of context access, tool integration, and quality gates
constitutes the harness at this stage without yet reaching agentic
operation.

Agentic usage patterns extend this by introducing systems that can generate or modify
artifacts, execute tests, and react to error messages. Such capabilities are found not only
in research prototypes but in production-ready tools such as GitHub Copilot~\cite{githubcloudagent},
Cursor Agent~\cite{cursoragent}, Claude Code~\cite{anthropicclaudecode}, and Amazon Q
Developer~\cite{amazon2024qdeveloper}. At this stage, the harness is extended with defined
execution environments, permission scoping, logging, and approval points. In practice, this
requires narrowly scoped permissions, defined sandboxes, and human approval for
production-proximate changes delivered as reviewable pull requests.

\subsection{Effort Shift and Economic Effects}

Generative AI redistributes rather than uniformly reduces effort in development projects.
Tasks such as boilerplate implementation, test drafting, and documentation can be
accelerated substantially. At the same time, quality assurance, verification, context
provisioning, and governance gain importance.

Economic impact is highly context-dependent. In standardized development tasks with clear
quality criteria, visible productivity gains are common. Studies with experienced developers
show, however, that review and correction overhead can exceed expected
gains~\cite{metr2025productivity,mozannar2024reading}. In safety-critical or complex legacy
environments, additional verification and coordination steps may consume a significant
portion of those gains. Beyond licensing and API costs, additional investment in training,
guidelines, security reviews, model approval, and process integration is required, especially
for agentic systems, where API-based models can generate highly variable runtime and token
costs requiring active cost control.

Table~\ref{tab:effort} summarizes typical qualitative effort shifts between conventional
and AI-driven development.

\begin{table}[!t]
\caption{Qualitative effort shift between conventional and AI-driven development}
\label{tab:effort}
\centering
\scriptsize
\renewcommand{\arraystretch}{1.3}
\setlength{\tabcolsep}{5pt}
\rowcolors{2}{gray!8}{white}

\begin{tabularx}{\textwidth}{L{3.0cm} Y Y Y}
\toprule
\textbf{Activity} &
\textbf{Conventional Effort} &
\textbf{AI-driven Effort} &
\textbf{Shift} \\
\midrule

Requirements Engineering &
Manual specification, coordination, and elaboration of requirements. &
AI assists with structuring, clarification, acceptance criteria, and initial drafts. &
Less writing effort; stronger focus on validation and domain alignment. \\

Implementation and Architecture Design &
Code and architecture variants predominantly created and adapted manually. &
AI supports architecture design variants, code generation, completion, refactoring, interfaces, and test preparation. &
Less routine implementation; greater effort for integration, review, and verification. \\

Test and Review &
Tests, reviews, and quality checks mostly manual or incomplete. &
AI generates tests, analyzes changes, and supports reviews. &
Faster artifact creation; additional effort for evaluation and quality assurance. \\

Deployment and Operations &
Deployment, error analysis, and incident communication heavily manual. &
AI supports build analysis, triage, log evaluation, and recovery preparation. &
Faster analysis and prioritization; greater need for control and approval processes. \\

Maintenance &
Legacy analysis, bug fixes, and documentation upkeep manual and time-intensive. &
AI assists with legacy analysis, change impact estimation, bug fixes, and documentation. &
Less analysis and onboarding effort; additional effort for quality assurance and documentation currency. \\

\bottomrule
\end{tabularx}
\end{table}

\subsection{Measuring Productivity}
\label{sec:metrics}

Productivity measurement in AI-driven development should not be
reduced to individual tools or isolated activity metrics. The number
of generated lines of code and raw suggestion acceptance rates say
little about whether throughput, quality, and stability actually
improve. Productivity assessment should therefore take multiple perspectives
into account, including speed, quality, operational stability,
security, human review overhead, and cost.

Metrics are instruments of measurement, not ends in themselves.
Practically useful are indicators that can be derived from existing
development and operations tooling without additional reporting
overhead. Established DevOps metrics, including DORA metrics, provide
a suitable starting point~\cite{dora2024stateofdevops}. Relevant
examples are lead time for changes, deployment frequency, change
failure rate, and recovery time.

In an AI-driven development context, these indicators can be
complemented by quality- and security-oriented measures such as test
coverage, security findings, frequency of human corrections, bugs
detected in regression tests, traceability of agentic changes, and
cost per agentic task. Table~\ref{tab:metric_explanations} in the
Appendix lists example metrics with data sources and their
significance in an AI-driven development context.

\section{Design Practices for Implementation}
\label{sec:gestaltung}

The transformation model describes increasing degrees of AI embedding. For practical application, however, this stage logic alone is not sufficient. Organizations must decide how project context is provisioned, how AI-generated artifacts are evaluated, how reviews are conducted, how tests are secured, and how agentic workflows are controlled. This section therefore translates the organizing framework into design practices that serve as a reference structure for the case study in Section~\ref{sec:caseA}.

The focus is on the organization of project knowledge, the technical safeguarding of AI-generated artifacts, and the governance of agentic workflows. Project knowledge must be made available in such a way that AI systems can work in a context-sensitive but controlled manner. AI-generated artifacts require reviews, tests, and technical quality gates. Agentic workflows additionally require mechanisms for limitation, logging, and human approval. The following design practices can therefore be understood as practical building blocks of a harness.

\subsection{Context Provisioning, Task Clarification, and Artifact Evaluation}

Integrated AI use requires reliably managed project context. AI systems should not merely
respond to isolated prompts but access curated, current information: source code,
architecture decisions, interface specifications, tickets, domain specifications, test
artifacts, operational documentation, and known incidents. What matters is not maximum
information quantity but targeted, traceable context selection. Context should be versioned,
assigned to responsible owners, and evaluated for currency, confidentiality, and relevance.
For larger codebases, repository indexing, retrieval mechanisms, and project-specific context
templates are needed to keep AI support reproducible and controllable.

Before implementation, AI can help structure requirements, surface open questions, and
propose acceptance criteria. Factual decisions must not, however, be delegated to the model.
A productive working pattern has AI generating suggestions for user stories, test cases,
interfaces, or non-functional requirements while product owners, architects, or domain
experts validate their correctness. The benefit lies in accelerated clarification and better
visibility of implicit assumptions, not in automated definition of business and technical goals.

AI-generated code should be treated like an external contribution that must be understood,
reviewed, and integrated. Before adoption, compilability, style conformance, architectural
compatibility, comprehensibility, testability, and security implications must be checked,
along with whether the proposal fits the existing codebase, uses dependencies correctly, and
avoids hallucinated content or unnecessary complexity. Responsibility for adoption remains
with the developers.

\subsection{Reviews, Tests, and Quality Gates}

In AI-integrated and AI-driven development, reviews must more sharply
separate generation from verification. AI can prepare pull requests,
explain changes, identify inconsistencies, and flag risks. Approval
should nonetheless be performed by humans, especially for
security-relevant, architectural, or production-proximate changes.

Practically useful review annotations make visible which parts of a
change were AI-assisted and what task description was used. They
should also document which tests were executed and what assumptions
the proposal contains.
This makes a review not only a code check but a check of the generation context.

AI can generate test cases, test data, and check scripts, but the existence of AI-generated
tests is not itself evidence of correctness. A particular risk arises when code and tests
originate from the same model context, since both may share similar assumptions or errors.
Tests should therefore be grounded in domain acceptance criteria, boundary cases, error
cases, and existing test requirements. AI-generated tests must also undergo review,
especially when they secure production-critical logic or replace existing tests.

For technical quality assurance, automated tests, static analysis,
security scans, dependency checks, and CI/CD pipelines remain central
quality gates. Tools such as GitHub Advanced Security, Snyk, Semgrep,
Trivy, and SonarQube address distinct aspects of this assurance,
including code scanning, dependency review, secret detection,
vulnerability analysis, license review, misconfiguration detection,
and quality rules
~\cite{githubadvancedsecurity,snyk2025docs,semgrep2025docs,trivy2025docs,sonarqube2025aicodeassurance}.
These mechanisms are especially important because AI-generated code
can appear plausible without being functionally correct, secure, or
maintainable.

\subsection{Tool Support and Agentic Workflows}

Modern AI development tools rely on several technical mechanisms.
They provide access to project context, use explicit project rules or
instruction files, operate in local or cloud-based environments, and
produce traceable artifacts such as diffs, pull requests, logs, or
test outputs.

For context access, repository search, semantic search, and explicit
context information are particularly relevant. Tools such as GitHub
Copilot and Cursor can retrieve relevant parts of a codebase through
text search, grep-like approaches, semantic retrieval, and agentic
search steps~\cite{githubcopilotdocs,cursoragent}. This avoids the
need to load entire codebases into a model window. Instead, relevant
files, symbols, documentation sections, or prior changes are retrieved
on demand and used as working context. Project knowledge must
therefore be findable, current, and sufficiently structured.

Search-based context can be complemented by explicit rule and
instruction files. Claude Code, for example, supports project-level
instructions and persistent context through files such as
\texttt{CLAUDE.md}~\cite{anthropicclaudecode}. Similar mechanisms can
take the form of repository rules, custom instructions, or
project-specific context templates. These mechanisms make
organizational knowledge more accessible to AI-assisted development
work. They can include architecture principles, coding conventions,
test requirements, security rules, domain terminology, and build or
deployment guidance.

Agentic tools extend this pattern by handling tasks in controlled
execution environments. GitHub Copilot Coding Agent can examine a
repository, create an implementation plan, make changes on a branch,
and prepare a pull request~\cite{githubcloudagent}. Cursor Agent,
Claude Code, and Amazon Q Developer also support agentic task
handling. Typical actions include file reading, code modification,
test execution, and code transformation
~\cite{cursoragent,anthropicclaudecode,amazon2024qdeveloper}.

Because of their extended reach, agentic tools must not operate
without constraints. Tasks should run in scoped branches, sandboxes,
or temporary environments. Scope, objective, permitted tools, cost
budget, and abort conditions should be defined in advance. Critical
actions require explicit human approval. This applies in particular
to merges, deployments, schema changes, permission changes, and
production-proximate configuration changes.

For auditability, relevant execution details should be logged. This
includes executed commands, models used, context sources, generated
artifacts, test runs, and costs. The combination of context access,
technical scoping, quality gates, and auditability helps make
agentic AI support governable in development organizations.

\subsection{Continuous Governance and Improvement}

Implementing the transformation model should be tightly coupled with measuring quality,
stability, review overhead, and costs. Organizations can start by establishing approved
tools and usage guidelines, then progressively integrate context systems, quality gates, and
review practices, restricting agentic workflows initially to clearly scoped tasks. Evaluation
should not be reduced to generated lines of code or usage frequency; more informative indicators are
cycle time, review overhead, defect density, security findings, human overrides, rollback
rates, cost per task, and audit coverage, as outlined in Section~\ref{sec:metrics} and
Table~\ref{tab:metric_explanations}.

AI-driven Software Development is thus not a one-time tool introduction but a continuous
learning and improvement process. New model capabilities, tools, security requirements, and
regulatory constraints must be regularly assessed and fed back into development practices,
standards, and governance mechanisms.

\section{Case Study: Mid-Sized Software Company}
\label{sec:caseA}

To ground the proposed transformation model, we consider a mid-sized software company that
develops custom software and client platforms, and takes end-to-end responsibility for
development, operations, maintenance, and modernization. Its business model is based
primarily on project-based development and consulting services. The company is responsible not only for development but also for
operating and maintaining complex client systems. The goal is to illustrate
how the framework applies to a typical development organization---assessing the recognizable
status quo and discussing possible development paths toward more AI-integrated processes.
The design practices from Section~\ref{sec:gestaltung} serve as implementation reference points.

\subsection{Initial Situation}

The organization is predominantly in an assistive usage phase. Tools such as Copilot,
Cursor, Claude, or ChatGPT are already in use, but typically at the level of individual
developers and projects, making adoption strongly dependent on particular teams and individuals. Differences are
particularly visible in tool selection, prompting practices, privacy assessment, and quality
assurance. Agile processes and DevOps orientation are established, and CI/CD pipelines are
used across projects. 

The weaker elements are shared governance mechanisms and organization-wide knowledge and
context structures. Project knowledge is distributed across repositories, ticket systems,
documentation, and operational artifacts. Unified review mechanisms for AI-generated
artifacts are largely absent. Existing tools, knowledge sources, and verification mechanisms
are not yet connected into a shared harness.

\subsection{Development Path}

The target is not fully autonomous software development. More realistic is a stepwise
deepening of generative AI integration into existing project activities, with responsibility
for architecture, quality assurance, governance, and domain decisions remaining with the
teams. 
Table~\ref{tab:target_ai_driven_company} summarizes possible changes between the
assistive starting state and a more AI-integrated development organization.

The proposed development path follows the transformation model described in this paper.
Initially, the focus is on the controlled use of existing AI tools. Building on this, AI can
gradually be embedded more deeply into development, testing, review, and knowledge processes.
Only at later stages do agentic forms of work become appropriate, for example for test
generation, pull-request preparation, or incident triage.

This development concerns not only tools but also organizational structures. Project-oriented
delivery is likely to be complemented by more platform- and service-oriented forms of work.
Shared guardrails and knowledge structures therefore gain importance.
\begin{table}[H]
\caption{Potential evolution from assistive to AI-integrated software development}
\label{tab:target_ai_driven_company}
\centering
\scriptsize
\renewcommand{\arraystretch}{1.12}
\setlength{\tabcolsep}{3pt}
\rowcolors{2}{gray!8}{white}
\begin{tabularx}{\textwidth}{L{2.5cm} Y Y}
\toprule
\textbf{Dimension} &
\textbf{Assistive Initial State} &
\textbf{Target State} \\
\midrule

Tool Support &
Individual use of isolated AI assistants. &
Approved AI tools with controlled context access. \\

Processes &
AI usage outside defined workflows. &
AI integrated into backlog, IDE, CI/CD, review, and operations. \\

Quality &
Manual quality checks applied after the fact. &
Automated quality, security, and compliance gates. \\

Knowledge &
Distributed project and operations knowledge. &
Versioned knowledge base accessible to teams and AI systems. \\

Roles &
Developers use AI individually. &
Teams orchestrate and review AI-generated artifacts. \\

Governance &
Individual decisions without consistent rules. &
Technical and organizational guardrails, governed permissions, and human approval mechanisms. \\

\bottomrule
\end{tabularx}
\end{table}

\subsection{Feasible Implementation Timeline}

A stepwise approach over approximately two years is realistic for a mid-sized software
company. Adoption should begin not with an organization-wide rollout but with one or two
reference projects
that are domain-relevant, technically manageable, and representative of recurring delivery
patterns. These projects enable tools, prompting and review practices, privacy boundaries,
test strategies, and initial metrics to be evaluated in a bounded context. Critically, the
reference projects should not only deliver local successes but be deliberately used to derive
reusable practices.

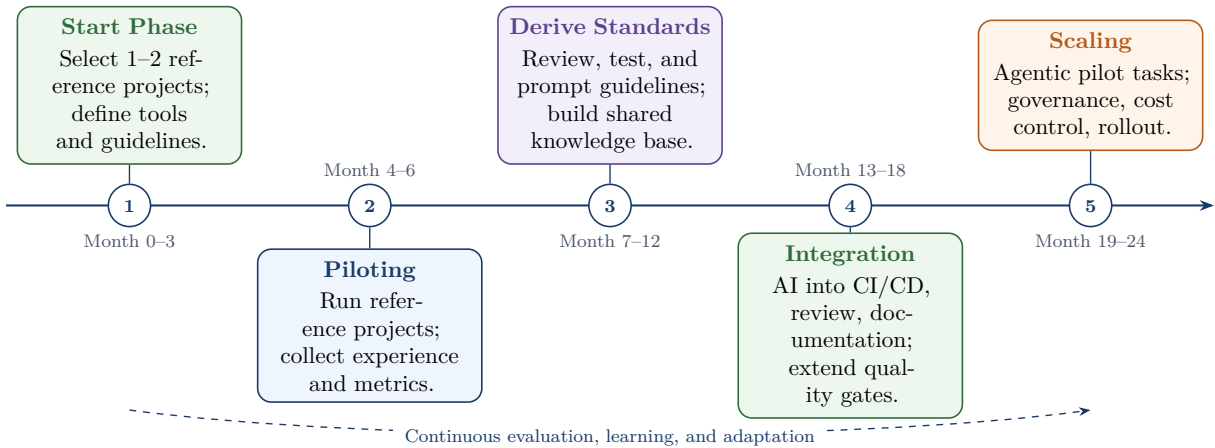
\begin{figure}[H]
\centering
\resizebox{\textwidth}{!}{%
\begin{tikzpicture}[
    font=\small,
    >=Stealth,
    phasebox/.style={
        rounded corners=6pt,
        line width=0.7pt,
        minimum width=3.45cm,
        minimum height=1.65cm,
        text width=3.10cm,
        align=center,
        inner sep=5pt,
        text=black
    },
    timepoint/.style={
        circle,
        fill=white,
        line width=0.7pt,
        minimum size=6.5mm,
        inner sep=0pt,
        font=\scriptsize\bfseries
    },
    tl/.style={
        line width=0.9pt,
        draw=aiBlueDark,
        -{Stealth[length=2.2mm]}
    },
    connector/.style={
        line width=0.6pt,
        draw=aiBlueDark
    },
    feedback/.style={
        dashed,
        line width=0.6pt,
        draw=aiBlueDark,
        -{Stealth[length=1.8mm]}
    },
    timelabel/.style={
        font=\scriptsize,
        text=aiGrayDark,
        align=center
    }
]

\draw[tl] (-9.3,0) -- (9.3,0);

\node[timepoint, draw=aiBlueDark, text=aiBlueDark] (p1) at (-7.4,0) {1};
\node[timelabel] at (-7.4,-0.55) {Month 0--3};

\node[timepoint, draw=aiBlueDark, text=aiBlueDark] (p2) at (-3.7,0) {2};
\node[timelabel] at (-3.7,0.55) {Month 4--6};

\node[timepoint, draw=aiBlueDark, text=aiBlueDark] (p3) at (0,0) {3};
\node[timelabel] at (0,-0.55) {Month 7--12};

\node[timepoint, draw=aiBlueDark, text=aiBlueDark] (p4) at (3.7,0) {4};
\node[timelabel] at (3.7,0.55) {Month 13--18};

\node[timepoint, draw=aiBlueDark, text=aiBlueDark] (p5) at (7.4,0) {5};
\node[timelabel] at (7.4,-0.55) {Month 19--24};

\node[
    phasebox,
    draw=aiGreenDark,
    fill=aiGreenFill
] (start) at (-7.4,1.85) {%
    \textbf{\textcolor{aiGreenDark}{Start Phase}}\\[2pt]
    Select 1--2 reference projects;\\
    define tools and guidelines.
};

\node[
    phasebox,
    draw=aiBlueDark,
    fill=aiBlueFill
] (pilot) at (-3.7,-1.85) {%
    \textbf{\textcolor{aiBlueDark}{Piloting}}\\[2pt]
    Run reference projects;\\
    collect experience and metrics.
};

\node[
    phasebox,
    draw=aiPurpleDark,
    fill=aiPurpleFill
] (std) at (0,1.85) {%
    \textbf{\textcolor{aiPurpleDark}{Derive Standards}}\\[2pt]
    Review, test, and prompt guidelines;\\
    build shared knowledge base.
};

\node[
    phasebox,
    draw=aiGreenDark,
    fill=aiGreenFill
] (int) at (3.7,-1.85) {%
    \textbf{\textcolor{aiGreenDark}{Integration}}\\[2pt]
    AI into CI/CD, review, documentation;\\
    extend quality gates.
};

\node[
    phasebox,
    draw=aiOrangeDark,
    fill=aiOrangeFill
] (scale) at (7.4,1.85) {%
    \textbf{\textcolor{aiOrangeDark}{Scaling}}\\[2pt]
    Agentic pilot tasks;\\
    governance, cost control, rollout.
};

\draw[connector] (p1.north) -- (start.south);
\draw[connector] (p2.south) -- (pilot.north);
\draw[connector] (p3.north) -- (std.south);
\draw[connector] (p4.south) -- (int.north);
\draw[connector] (p5.north) -- (scale.south);

\draw[feedback]
    (-7.4,-3.15) .. controls (-3.7,-3.65) and (3.7,-3.65) .. (7.4,-3.15);

\node[
    font=\scriptsize,
    text=aiBlueDark,
    fill=white,
    inner xsep=5pt,
    inner ysep=1pt
] at (0,-3.58) {Continuous evaluation, learning, and adaptation};

\end{tikzpicture}}
\caption{Proposed implementation timeline for stepwise adoption of AI-driven Software
Development in a mid-sized software company.}
\label{fig:case_timeline}
\end{figure}

In a first phase, tool approval, role clarification, and reference project selection take
precedence. Experience from these projects is then systematically evaluated to derive
organizational standards. Broader integration into CI/CD pipelines, knowledge platforms,
and operational processes is only sensible on this basis. Toward the end of a two-year
period, initial agentic workflows can be piloted for clearly scoped tasks and then
progressively scaled. Figure~\ref{fig:case_timeline} sketches this timeline.

The timeline is deliberately a guiding framework. Depending on project portfolio,
client proximity, and existing platform maturity, individual steps may proceed faster or
slower. What matters is not the exact duration but the sequence: targeted piloting first,
then standardization, then broader process integration, and only then controlled scaling of
agentic workflows.

\subsection{Implementation Along the Design Practices}

The implementation follows the practices outlined in
Section~\ref{sec:gestaltung}. The existing tool landscape is not
replaced, but extended with controlled AI capabilities. The transition
therefore does not consist of deploying a single AI tool. It consists
of progressively building a harness that connects context
provisioning, tool use, quality assurance, and governance.

In the early phases, the organization should define shared guidelines
for data privacy, prompting, review, and tool approval. These
guidelines establish a common baseline for teams that already use AI
tools in different ways. They also help clarify which data may be
used, which tools are approved, and how AI-generated artifacts should
be reviewed before they enter shared repositories or client-facing
deliverables.

As implementation progresses, shared knowledge and context systems
become more important. Repository knowledge, documentation, tickets,
and operational data should be made accessible in a structured and
maintainable form. This does not mean that all information must be
loaded into AI systems. Rather, relevant context should be retrievable
when needed and kept sufficiently current. On this basis, AI systems
can support requirements analysis, test generation, code review,
documentation, and selected operational tasks more systematically.

For a concrete codebase, such a harness could consist of several
connected elements. The repository would expose relevant source files,
architecture notes, coding conventions, test suites, issue references,
and deployment information as retrievable context. Project-specific
instruction files could describe architectural boundaries, naming
rules, security constraints, and expectations for tests and
documentation. Current tools increasingly support such mechanisms, for
example through project-level instruction files, repository rules, or
tool-specific context files
~\cite{anthropicclaudecode,cursoragent,githubcopilotdocs}. AI tools
would then operate against this structured context rather than against
isolated prompts. A proposed change would be created on a separate
branch, accompanied by a short rationale, affected files, executed
tests, and known assumptions. Before integration, the change would
pass through automated tests, static analysis, dependency checks, and
human review. An illustrative example of how such a repository structure can make
the harness visible in a codebase is provided in
\hyperref[app:codebase_harness]{Appendix~B}. The example should be understood as a repository-level
concretization. It mainly shows the part of a harness that becomes
visible in the codebase. A complete harness also includes the
execution environment, context selection, tool access, permission
boundaries, quality gates, logging, cost control, and human approvals.
The codebase represents only one part of the technical and
organizational control environment.

Artifact evaluation should be integrated more tightly with the
development workflow. Reviews, automated tests, security checks, and
dependency verification should not be treated as separate downstream
activities. They should become part of the normal path through which
AI-assisted changes are accepted, rejected, or revised. This is
especially important when AI-generated artifacts affect architecture,
security, maintainability, or production-related behavior.

Agentic tools are most appropriate where tasks are clearly scoped and
technically controllable. Examples include pull-request preparation,
test execution, and operational event analysis. Even in these cases,
the agentic workflow should remain bounded by explicit objectives,
limited permissions, traceable changes, and documented results.
Automated checks, audit logs, and human approval for
production-proximate actions form the control framework that makes
such use practically viable.

\subsection{Effort Shift and Organizational Implications}

With increasing AI support, development effort is redistributed. Routine tasks---boilerplate
code, straightforward documentation, standardized test cases---can be partly automated.
At the same time, demands on review, context maintenance, architecture work, and quality
assurance increase. The value contribution shifts from artifact creation toward verification
and accountability for sound solutions.

This has economic implications. Conventional time-and-material models, heavily oriented
toward billable implementation time, become less meaningful when part of creation is
accelerated or automated. The economic emphasis shifts toward delivery capability, quality,
stability, and continuous product ownership---opening space for more service- and
platform-oriented offerings such as AI-assisted modernization programs or managed AI
development services.

Role profiles also change. Implementation remains important but is increasingly complemented
by AI engineering, platform operations, knowledge management, and governance. Technical
system understanding, security competence, and the ability to evaluate AI-generated artifacts
gain in importance.

\subsection{Risks and Productivity Measurement}

Deeper AI integration introduces new risks, including insecure generated code and opaque
agentic changes. Organizational guidelines alone cannot fully address these risks. Technical
control mechanisms embedded in the development and operations environment are therefore
essential. Governance becomes part of the workflow rather than a separate overlay.

For observing and evaluating the transformation, a small set of automatically collectable
metrics is initially sufficient. Benefits should not be expected only after a complete
transition to AI-driven development. Even AI-integrated stages can make improvements visible,
for example through faster preparation of reviews, more systematic test generation, better
documentation support, or earlier detection of defects and security findings. In early
phases, delivery metrics such as lead time, deployment frequency, and failure rate are useful
in combination with defect density, test coverage, and vulnerability rate. As AI integration
deepens, additional indicators such as review acceptance rate and regression detection rate
provide further signal.

\subsection{Summary Assessment of the Case Study}

The case study shows that the transition to AI-driven Software
Development in a medium-sized software company should proceed
step by step and based on practical experience. A realistic path
begins with a small number of reference projects. The experiences
gained from these projects form the basis for company-wide standards
and broader integration into the development process.

The company considered already has a comparatively good starting
position. In organizations with less standardized processes, weaker
test automation, or fragmented knowledge structures, the development
of these foundations would have to be given greater priority.
After the introduction phase, AI is no longer merely an additional
tool used by individual developers, but an integral part of software
development. At the same time, it remains embedded in technical and
organizational control mechanisms. 

The process should not be
understood as a one-time transformation, but must be continuously
adapted to new AI-related insights, model capabilities, tools, and
regulatory requirements.
A balance between standardization and learning capability is important.
Quick wins can make benefits visible at an early stage, while controlled
spaces for experimentation ensure that new tools and methods are
continuously evaluated and, where suitable, transferred into shared
standards.
The metrics described support this continuous evaluation and
improvement.

\clearpage
\section{Conclusion}
\label{sec:conclusion}

\subsection{Summary}

This paper has developed a pragmatic framework for AI-driven Software Development and
grounded it in an exploratory case study. The central contribution is not the replacement
of classical software development but a structured account of how generative AI can be
systematically embedded in development processes. The framework distinguishes assistive,
integrated, and agentic usage forms---not as a rigid maturity model but as an organizing
logic for practitioners.

The harness concept describes the operative embedding of AI systems across context
provisioning, tool use, verification, governance, and human approval. The literature review
and practice show that generative AI now affects diverse activities across the software
lifecycle beyond code generation, while effects and risks remain strongly context-dependent.
Productivity gains arise primarily where AI is embedded in stable, well-organized development
processes. Without those foundations, the benefits are limited and the risks heightened.

The design practices derived in Section~\ref{sec:gestaltung} concretize the transition for
context provisioning, task clarification, artifact evaluation, reviews, tests, agentic
workflows, and measurement. They serve simultaneously as building blocks of a harness. The
case study illustrates that AI integration should be understood as a continuous learning and
improvement process, not a one-time transformation. Metrics serve not only to measure
success but to continuously evaluate benefit, quality, review overhead, risk, and cost.

\subsection{Limitations}

The proposed transformation model describes only one possible path toward AI-driven Software
Development, and the design practices are heuristic concretizations rather than empirically
validated methods. The comparison of related approaches demonstrates overlaps and
distinctions but does not substitute for empirical validation of the proposed transition
logic.

The work rests on literature analysis, practitioner reports, the authors' project
experience, and one exploratory case study. It therefore does not allow statistical generalization. Parts of
the analysis rely on publicly available information; internal architecture, cost, or
governance data were not fully accessible. The tool landscape is also evolving rapidly, so
specific tools and examples represent a snapshot.

\subsection{Outlook}

Future work should prioritize empirical investigation. This includes
longitudinal studies on the effects of AI assisted development,
comparisons of integration strategies, and research on governance,
role change, and the economics of agentic systems. Open questions also
remain around the evaluation of agentic development environments, the
management of external model providers, and the longer term
integration of context and knowledge systems into existing platforms.

This research area is likely to evolve rapidly. Recent work on harness
engineering indicates that the surrounding runtime environment, tool
access, project memory, observability, verification, and permission
structures are becoming central to the performance and governability of
coding agents~\cite{openai2026harness,zhong2026aiharness}. Emerging
approaches such as Retrospective Harness Optimization suggest that
future systems may not only use a predefined harness, but continuously
adapt it based on previous execution trajectories and observed failure
patterns~\cite{pan2026rho}. It can therefore be assumed that these
developments will also affect the structure of the software development
process itself. Activities such as task decomposition, review,
verification, documentation, and operational analysis may become more
closely connected to agentic execution environments and their harness
structures.

A central question is how such harness structures can be designed,
measured, and evolved in real development organizations without
weakening human accountability, quality assurance, and organizational
control.

\section*{Disclosure of Generative AI Use}
\addcontentsline{toc}{section}{Disclosure of Generative AI Use}

Generative AI tools were used in an assistive capacity for language revision, structuring,
and the creation of figures based on the authors' specifications. Conceptual design,
verification, and responsibility for the content of this paper rest with the authors.

\clearpage
\appendix

\section*{A\quad Supplementary Overviews}
\addcontentsline{toc}{section}{A\quad Supplementary Overviews}

This appendix complements the main text with four overviews.
Table~\ref{tab:terminology_ai_driven} classifies related
terms and selected sources around the working term
AI-driven Software Development.
Table~\ref{tab:ai_capabilities_by_activity} describes
AI capabilities across central project activities and the
control mechanisms required for them.
Table~\ref{tab:related_evidence_risks} condenses central
findings from the literature on productivity, security,
adoption, and agentic systems.
Table~\ref{tab:metric_explanations} explains example
metrics for delivery capability, quality, governance, and
agentic development processes.

The operationalization of these metrics depends strongly on the
tool chain used, the organizational structure, and the degree of
automation. Many metrics should be understood as proxy indicators
and should not be interpreted in isolation. High acceptance rates
of AI-generated artifacts or low levels of human intervention, for
example, do not necessarily imply higher quality or lower risk.

\newcolumntype{Z}{>{\justifying\arraybackslash}X}
\setlength{\parindent}{0pt}
\begin{table}[!htbp]
\caption{Conceptual classification of related concepts in AI-driven Software Development}
\label{tab:terminology_ai_driven}
\centering
\scriptsize
\renewcommand{\arraystretch}{1.18}
\setlength{\tabcolsep}{4pt}
\rowcolors{2}{gray!8}{white}

\begin{tabularx}{\textwidth}{L{3.3cm} Z L{3.2cm}}
\toprule
\textbf{Term} &
\multicolumn{1}{l}{\textbf{Meaning in this paper}} &
\textbf{Example sources} \tabularnewline
\midrule

AI-driven Software Development &
\noindent Umbrella working term for AI-supported, extended, or partially automated activities across the Software Development Lifecycle. &
\cite{terragni2025future,hassan2024ainative} \tabularnewline

AI-assisted Software Engineering &
\noindent Assistive form in which AI supports individual activities such as implementation, review, testing, documentation, or explanation; frequent starting point of the transformation. &
\cite{barke2023groundedcopilot,liang2024usability,sergeyuk2024practice} \tabularnewline

LLM4SE &
\noindent Technology- and research-field perspective on the application of large language models in software engineering. &
\cite{hou2024llmse,fan2023survey} \tabularnewline

Generative software development &
\noindent Artifact-oriented perspective on the generation of development artifacts, classically model- or domain-specific language-based and currently increasingly LLM-based. &
\cite{rumpe2010generative,jiang2024codegeneration,zhang2024agilegen} \tabularnewline

AI-native / AI-first Software Engineering &
\noindent Transformative or strategic perspective on a more fundamental realignment of development processes, roles, and toolchains around AI capabilities. &
\cite{hassan2024ainative,li2025aiteammates} \tabularnewline

Agentic Software Engineering &
\noindent Agent-based form in which AI systems plan tasks, use tools, and iteratively improve results; relevant for autonomous or semi-autonomous workflows. &
\cite{yang2024sweagent,hong2024metagpt,wu2024autogen,wang2024openhands} \tabularnewline

\bottomrule
\end{tabularx}
\end{table}

\FloatBarrier

\begin{table}[!htbp]
\caption{AI-supported capabilities and control needs across central project activities}
\label{tab:ai_capabilities_by_activity}
\centering
\scriptsize
\renewcommand{\arraystretch}{1.12}
\setlength{\tabcolsep}{4pt}
\rowcolors{2}{gray!8}{white}

\begin{tabularx}{\textwidth}{L{2.8cm} Z Z}
\toprule
\textbf{Activity} &
\multicolumn{1}{l}{\textbf{AI-supported capability}} &
\multicolumn{1}{l}{\textbf{Control need}} \tabularnewline
\midrule

Requirements Engineering &
\noindent Structuring requirements, deriving user stories and acceptance criteria. AI is particularly useful here for condensation, variant generation, and making implicit assumptions visible. &
\noindent Domain legitimacy, prioritization, and responsibility for requirements remain with the product owner and domain stakeholders. \tabularnewline

Architecture Design and Implementation &
\noindent Sketching alternative architecture designs, generating complete code blocks, code completion, refactoring, adapter logic, bug fixing, and changes across multiple files. This is particularly effective for recurring development tasks and when working with unfamiliar frameworks. &
\noindent Architecture responsibility, integration into build processes, tests, and security requirements remain central; additional review and verification effort must be bounded. \tabularnewline

Test and Review &
\noindent Suggestions for unit tests, test data, edge cases, test scenarios, and pull-request summaries. AI can make test gaps visible earlier, accelerate checking tasks, and prepare reviews. &
\noindent AI-generated verification artifacts must be executed, evaluated, and secured through human review. \tabularnewline

Deployment and Operations &
\noindent Support for deployment pipelines, build failure analysis, release notes, log summaries, incident classification, and drafts for operational runbooks. AI is particularly useful for operational analysis, prioritization, and preparation of recovery or rollback measures. &
\noindent Production interventions require permission scoping, audit logs, and approval processes. \tabularnewline

Maintenance &
\noindent Analysis of legacy code, visibility into technical debt, estimation of change impact, and support for bug fixes, refactorings, and technical documentation. &
\noindent Changes should remain small, test-secured, and reversible; documentation must be reconciled with the actual system state. \tabularnewline

\bottomrule
\end{tabularx}
\end{table}

\FloatBarrier
\

\begingroup
\scriptsize
\renewcommand{\arraystretch}{1.12}
\setlength{\tabcolsep}{4pt}

\begin{longtable}{>{\raggedright\arraybackslash}p{3.2cm}>{\centering\arraybackslash}p{0.9cm}>{\setlength{\parindent}{0pt}\justifying\arraybackslash}p{\dimexpr\textwidth-3.2cm-0.9cm-6\tabcolsep\relax}}
\caption{Selected research and practice sources on AI-driven Software Development}
\label{tab:related_evidence_risks}\tabularnewline

\toprule
\rowcolor{gray!18}
\textbf{Work} &
\textbf{Cat.} &
\multicolumn{1}{l}{\textbf{Classification and relevance}} \tabularnewline
\midrule
\endfirsthead

\caption{Selected research and practice sources on AI-driven Software Development (continued)}\tabularnewline

\toprule
\rowcolor{gray!18}
\textbf{Work} &
\textbf{Cat.} &
\multicolumn{1}{l}{\textbf{Classification and relevance}} \tabularnewline
\midrule
\endhead

\midrule
& &
\makebox[\linewidth][r]{\emph{Continued on the next page}}\tabularnewline
\endfoot

\bottomrule
\endlastfoot

Hou et al.~\cite{hou2024llmse} &
beg &
\noindent Broad literature review on applications of large language models in software engineering; motivates the lifecycle-wide view of AI-supported development. \tabularnewline

\rowcolor{gray!8}
Fan et al.~\cite{fan2023survey} &
beg &
\noindent Shows hallucinations, verification, and evaluation as central problems and supports the importance of quality gates. \tabularnewline

Hassan et al.~\cite{hassan2024ainative} &
beg &
\noindent Classifies AI-native Software Engineering as intent-oriented collaboration with AI teammates and supports the transformative perspective. \tabularnewline

\rowcolor{gray!8}
Rumpe et al.~\cite{rumpe2010generative} &
beg &
\noindent Clarifies the distinction between classical model- or domain-specific language-based artifact generation and LLM-based generation. \tabularnewline

Chen et al.~\cite{chen2021codex} &
mod &
\noindent Present Codex as an early large language model for code generation and provide a foundation for the technical development of modern coding assistants. \tabularnewline

\rowcolor{gray!8}
Roziere et al.~\cite{roziere2023codellama} &
mod &
\noindent Describe Code Llama as a specialized model family for programming tasks and strengthen the evidence on domain-specifically trained code models. \tabularnewline

Li et al.~\cite{li2023starcoder} &
mod &
\noindent Introduce StarCoder as an open code model and support comparability and reproducibility in research on AI-based code generation. \tabularnewline

\rowcolor{gray!8}
Guo et al.~\cite{guo2024deepseekcoder} &
mod &
\noindent Present DeepSeek-Coder as a powerful code model and extend the technical foundation for code generation, code understanding, and repair. \tabularnewline

Hui et al.~\cite{hui2024qwen25coder} &
mod &
\noindent Describe Qwen2.5-Coder as a specialized model family for programming tasks and show progress in model performance and code understanding. \tabularnewline

\rowcolor{gray!8}
Mistral AI~\cite{mistral2024codestral} &
mod &
\noindent Presents Codestral as a code-oriented language model and complements the model landscape for AI-supported software development. \tabularnewline

OpenAI~\cite{openai2025gpt41} &
mod &
\noindent Describes GPT-4.1 as a powerful model with relevance for programming, context, and tool-use tasks. \tabularnewline

\rowcolor{gray!8}
OpenAI~\cite{openai2025gpt5} &
mod &
\noindent Describes GPT-5 as a current model generation with extended capabilities for programming, tool use, and agentic forms of work. \tabularnewline

Google~\cite{google2025gemini25pro} &
mod &
\noindent Presents Gemini 2.5 Pro as a powerful model with relevance for complex development, context, and reasoning tasks. \tabularnewline

\rowcolor{gray!8}
Anthropic~\cite{anthropic2025claude4} &
mod &
\noindent Describes Claude 4 as a model generation with relevance for programming, agent behavior, and long-context processing. \tabularnewline

Jiang et al.~\cite{jiang2024codegeneration} &
mod &
\noindent Provides an overview of code generation with language models and classifies code generation as a central research field. \tabularnewline

\rowcolor{gray!8}
OpenAI~\cite{openai2026harness} &
har &
\noindent Motivates harness engineering as a design task for agentic coding systems and emphasizes runtime environment, tool access, context, and control. \tabularnewline

Böckeler~\cite{bockeler2026harness} &
har &
\noindent Classifies harness engineering from the perspective of practical coding-agent use and shows the importance of technical and organizational embedding. \tabularnewline

\rowcolor{gray!8}
Zhong and Zhu~\cite{zhong2026aiharness} &
har &
\noindent Conceptualize AI Harness Engineering as a runtime substrate for foundation-model software agents and strengthen the technical grounding of the harness concept. \tabularnewline

Amazon~\cite{amazon2024qdeveloper} &
har &
\noindent Describes Amazon Q Developer as a product-oriented development environment and shows the integration of AI support into developer workflows. \tabularnewline

\rowcolor{gray!8}
Anthropic~\cite{anthropicclaudecode} &
har &
\noindent Describes Claude Code as an agentic development tool and strengthens the relevance of tool use, repository context, and developer control. \tabularnewline

Cursor~\cite{cursoragent} &
har &
\noindent Provides agentic functions in an AI-oriented development environment and shows the shift from assistance toward more integrated workflows. \tabularnewline

\rowcolor{gray!8}
GitHub~\cite{githubcloudagent} &
har &
\noindent Describes cloud-agent functions for development tasks and complements the discussion of outsourced, agentic processing of software changes. \tabularnewline

GitHub~\cite{githubcopilotdocs} &
har &
\noindent Documents Copilot functions and shows the product-oriented embedding of AI support in IDEs, repositories, and development processes. \tabularnewline

\rowcolor{gray!8}
AWS~\cite{aws2025addlc} &
ver &
\noindent Classifies AI-supported development as an AI-Driven Development Life Cycle and supports the lifecycle-wide perspective of the paper. \tabularnewline

DORA~\cite{dora2025aicapabilities} &
ver &
\noindent Describes organizational capabilities for effective AI support and supports the role of platform maturity, data access, and clear guardrails. \tabularnewline

\rowcolor{gray!8}
Forrester~\cite{forrester2024turingbots} &
ver &
\noindent Classifies TuringBots as a development line of AI-supported software development and illustrates changes in tool landscapes and developer roles. \tabularnewline

Gartner~\cite{gartner2025softwaretrends} &
ver &
\noindent Describes current trends in software development and supports the classification of AI as a strategic influence on development organizations. \tabularnewline

\rowcolor{gray!8}
McKinsey~\cite{mckinsey2025softwarevalue} &
ver &
\noindent Emphasizes economic potential of generative AI in software development while also pointing to the importance of organizational implementation. \tabularnewline

Thoughtworks~\cite{thoughtworks2024aitools} &
ver &
\noindent Classifies AI tools in modern software development from a practice-oriented perspective and supports the view of AI as an integrated part of development platforms. \tabularnewline

\rowcolor{gray!8}
Zhang et al.~\cite{zhang2024agilegen} &
ver &
\noindent Connects generative software development with human-AI teamwork and testable requirements and strengthens the link to requirements engineering and validation. \tabularnewline

Zadenoori et al.~\cite{zadenoori2025requirements} &
ver &
\noindent Shows support for elicitation, validation, and specification analysis and demonstrates relevance for requirements engineering. \tabularnewline

\rowcolor{gray!8}
Norheim et al.~\cite{norheim2024llmre} &
ver &
\noindent Analyze challenges in using large language models in requirements engineering, especially reproducibility, evaluation, and domain adaptation. \tabularnewline

Mavin et al.~\cite{mavin2009ears} &
ver &
\noindent Provide EARS as an approach for more precise, consistent, and testable natural-language requirements. \tabularnewline

\rowcolor{gray!8}
Fischbach et al.~\cite{fischbach2022cira} &
ver &
\noindent Show how acceptance tests can be derived from requirements and support the mutual use of requirements and tests. \tabularnewline

Wang et al.~\cite{wang2019acceptancetests} &
ver &
\noindent Investigate the automatic generation of acceptance test cases from use-case specifications and strengthen the connection between specification and testing. \tabularnewline

\rowcolor{gray!8}
Dau et al.~\cite{xmainframe2024} &
ver &
\noindent Address mainframe modernization with specialized language models and show the relevance of AI for legacy analysis and transformation tasks. \tabularnewline

Kumar et al.~\cite{ibm2024coboljava} &
ver &
\noindent Investigate the automated validation of COBOL-to-Java transformations and emphasize verifiable migration artifacts. \tabularnewline

\rowcolor{gray!8}
Gandhi et al.~\cite{gandhi2024coboljava} &
ver &
\noindent Address the translation of COBOL to Java and illustrate opportunities and limits of AI-supported legacy migration. \tabularnewline

Solovyeva et al.~\cite{solovyeva2025plsqljava} &
ver &
\noindent Analyze PL/SQL-to-Java transformation as a case study of automated legacy-code translation and extend the evidence on AI-supported modernization. \tabularnewline

\rowcolor{gray!8}
Terragni et al.~\cite{terragni2025future} &
mki &
\noindent Describes software engineering as an increasingly collaborative human-AI process and supports the organizational perspective of the transformation model. \tabularnewline

Treude~\cite{treude2025taxonomy} &
mki &
\noindent Structures AI-supported software engineering conceptually and supports the terminological classification of different forms of support. \tabularnewline

\rowcolor{gray!8}
Li et al.~\cite{li2025aiteammates} &
mki &
\noindent Describe AI systems as possible teammates in software development and strengthen the perspective on role change and collaboration. \tabularnewline

Hamza et al.~\cite{hamza2023humanai} &
mki &
\noindent Investigate human-AI collaboration in software engineering and complement the empirical perspective on acceptance, control, and division of work. \tabularnewline

\rowcolor{gray!8}
Liang et al.~\cite{liang2024usability} &
mki &
\noindent Emphasizes usability and trust as acceptance factors and shows that tool quality alone is not sufficient. \tabularnewline

Sergeyuk et al.~\cite{sergeyuk2024practice} &
mki &
\noindent Documents different usage patterns across teams and projects and supports the need for processual embedding. \tabularnewline

\rowcolor{gray!8}
Barke et al.~\cite{barke2023groundedcopilot} &
mki &
\noindent Distinguishes exploratory and accelerating use and highlights different control needs. \tabularnewline

Vaithilingam et al.~\cite{vaithilingam2022expectation} &
mki &
\noindent Shows differences between expectations and actual usage experience; adoption depends strongly on practical usability. \tabularnewline

\rowcolor{gray!8}
Mozannar et al.~\cite{mozannar2024reading} &
mki &
\noindent Describes cognitive review effort for AI-generated suggestions and makes productivity measurement beyond raw suggestion acceptance necessary. \tabularnewline

Russo~\cite{russo2024adoption} &
mki &
\noindent Emphasizes organizational factors in the adoption of generative AI and shows transformation as more than a tooling issue. \tabularnewline

\rowcolor{gray!8}
Khemka and Houck~\cite{khemka2024support} &
mki &
\noindent Identifies trust, transparency, and control as central factors and connects acceptance with governance. \tabularnewline

Peng et al.~\cite{peng2023copilot} &
pro &
\noindent Reports substantial time gains in programming tasks with GitHub Copilot and supports assistive use in well-bounded tasks. \tabularnewline

\rowcolor{gray!8}
Cui et al.~\cite{cui2025highskilled} &
pro &
\noindent Shows context-dependent productivity effects and underlines the importance of process and organizational context. \tabularnewline

METR~\cite{metr2025productivity} &
pro &
\noindent Shows longer task completion times for experienced developers under AI support and points to review and integration effort. \tabularnewline

\rowcolor{gray!8}
DORA~\cite{dora2024stateofdevops} &
pro &
\noindent Connects software delivery performance, organizational practices, and AI use and provides an empirical reference point for transformation capability. \tabularnewline

Hashimoto~\cite{hashimoto2026aiadoption} &
pro &
\noindent Describes the stepwise adoption of AI tools in real development practices and supports the perspective of incremental adoption. \tabularnewline

\rowcolor{gray!8}
GitHub~\cite{github2024accenture} &
pro &
\noindent Documents practical experience on AI use in large development organizations and supports the view of context-dependent productivity effects. \tabularnewline

GitHub~\cite{github2024codequality} &
pro &
\noindent Reports relationships between AI support and code quality and complements the discussion of productivity- and quality-related effects. \tabularnewline

\rowcolor{gray!8}
Google~\cite{google2024q3} &
pro &
\noindent Provides practical evidence for the use of AI in development processes and points to organizational scaling effects. \tabularnewline

Shopify~\cite{shopify2025memo} &
pro &
\noindent Documents organizational expectations regarding AI use and shows that AI adoption also touches leadership, governance, and culture. \tabularnewline

\rowcolor{gray!8}
The Verge~\cite{theverge2025shopify} &
pro &
\noindent Reports on Shopify's AI-related organizational practice and complements the discussion of more mandatory forms of AI adoption. \tabularnewline

Pearce et al.~\cite{pearce2022asleep} &
sec &
\noindent Finds vulnerable suggestions in security-critical scenarios and motivates security checks and approvals. \tabularnewline

\rowcolor{gray!8}
Sandoval et al.~\cite{sandoval2023lostatc} &
sec &
\noindent Shows that LLM assistants can influence security decisions and that security risks can already emerge early. \tabularnewline

Khoury et al.~\cite{khoury2023securechatgpt} &
sec &
\noindent Shows that plausible AI-generated code can be insecure and that code generation does not replace security review. \tabularnewline

\rowcolor{gray!8}
Liu et al.~\cite{liu2024evalplus} &
sec &
\noindent Shows that classical benchmarks can overestimate model quality and that project-specific tests remain necessary. \tabularnewline

Ouyang et al.~\cite{ouyang2024nondeterminism} &
sec &
\noindent Describes non-determinism as a challenge for reproducibility, verification, and controllability of generative systems. \tabularnewline

\rowcolor{gray!8}
Spracklen et al.~\cite{spracklen2024packagehallucinations} &
sec &
\noindent Analyzes hallucinated package names as a supply-chain risk and strengthens the importance of dependency checking. \tabularnewline

Zhang et al.~\cite{zhang2025providerbias} &
sec &
\noindent Shows systematic differences between model providers and points to the importance of reflected tool choice. \tabularnewline

\rowcolor{gray!8}
GitHub~\cite{githubadvancedsecurity} &
sec &
\noindent Describes Advanced Security as a mechanism for code scanning, secret detection, and dependency checking and supports technical quality gates. \tabularnewline

Semgrep~\cite{semgrep2025docs} &
sec &
\noindent Documents static analysis and rule-based code checking and supports the safeguarding of AI-generated code artifacts. \tabularnewline

\rowcolor{gray!8}
Snyk~\cite{snyk2025docs} &
sec &
\noindent Documents security checks for dependencies and code and strengthens the importance of automated security controls. \tabularnewline

SonarQube~\cite{sonarqube2025aicodeassurance} &
sec &
\noindent Describes AI Code Assurance and supports quality assurance for AI-generated code contributions. \tabularnewline

\rowcolor{gray!8}
Trivy~\cite{trivy2025docs} &
sec &
\noindent Documents container, dependency, and infrastructure scanning and complements the technical safeguarding of build and deployment pipelines. \tabularnewline

Chen et al.~\cite{chen2024codeevaluation} &
eva &
\noindent Compares evaluation methods for AI-based code generation and shows the importance of suitable metrics and benchmarks. \tabularnewline

\rowcolor{gray!8}
Xia et al.~\cite{xia2023apr} &
eva &
\noindent Shows strong results of large language models in automated program repair and points to increasing AI-based bug fixing. \tabularnewline

Jimenez et al.~\cite{jimenez2024swebench} &
eva &
\noindent Introduces SWE-bench with more realistic development tasks and underlines the importance of project-related evaluation. \tabularnewline

\rowcolor{gray!8}
Yang et al.~\cite{yang2024sweagent} &
eva &
\noindent Shows how language models can use files, tests, and development tools and extends classical assistance models toward agents. \tabularnewline

Hong et al.~\cite{hong2024metagpt} &
eva &
\noindent Describes coordinated agent roles for development tasks and structures more complex agentic development processes. \tabularnewline

\rowcolor{gray!8}
Wu et al.~\cite{wu2024autogen} &
eva &
\noindent Shows agent communication and coordination through conversational structures and extends agentic development environments. \tabularnewline

Wang et al.~\cite{wang2024openhands} &
eva &
\noindent Provides an open platform for software development agents and strengthens reproducible agent environments. \tabularnewline

\rowcolor{gray!8}
Zhang et al.~\cite{zhang2024agents} &
eva &
\noindent Provides an overview of autonomous LLM agents and classifies agentic systems as an independent research field. \tabularnewline

Sethupathy~\cite{sethupathy2023zerotouchdevops} &
eva &
\noindent Describes Zero-Touch DevOps as a far-reaching automation approach and provides a reference point for agentic process automation. \tabularnewline

\end{longtable}

\vspace{0.5em}
\noindent\footnotesize\textbf{Legend:}
beg = conceptual classification of AI-driven Software Development;
mod = LLMs and code models in software engineering;
har = harness, tool integration, and development environments;
ver = related approaches to AI-supported software development;
mki = human-AI collaboration;
pro = productivity and empirical impact;
sec = security, correctness, and supply-chain risks;
eva = project-related evaluation and agentic systems.

\endgroup

\FloatBarrier
\clearpage

\begin{table}[H]
\caption{Example metrics for measuring productivity, quality, and governance in AI-driven Software Development}
\label{tab:metric_explanations}
\centering
\scriptsize
\renewcommand{\arraystretch}{1.12}
\setlength{\tabcolsep}{3pt}
\rowcolors{2}{gray!8}{white}

\begin{tabularx}{\textwidth}{L{3.0cm} L{3.2cm} Z Z}
\toprule
\textbf{Metric} &
\textbf{Source} &
\textbf{Meaning} &
\textbf{Operationalization} \tabularnewline
\midrule

Lead Time for Changes &
Git, CI/CD &
\noindent Time from change to production. &
\noindent Median from merge to successful production deployment per PR. \tabularnewline

Deployment Frequency &
CI/CD, deployment platform &
\noindent Frequency of production deployments. &
\noindent Deployments per day, week, or sprint. \tabularnewline

Change Failure Rate &
Incident and deployment systems &
\noindent Share of failed deployments. &
\noindent Deployments with incident, rollback, or hotfix. \tabularnewline

Defect Density &
Issues, tests &
\noindent Defects relative to change size. &
\noindent Defects per 1{,}000 changed lines of code or release. \tabularnewline

Test Coverage &
Test and coverage tools &
\noindent Degree of automated test coverage. &
\noindent Covered lines or code paths in percent. \tabularnewline

Vulnerability Rate &
SAST, SCA, security tools &
\noindent Detected vulnerabilities. &
\noindent New findings per sprint, release, or 1{,}000 lines of code. \tabularnewline

Review Acceptance Rate &
Pull-request platforms &
\noindent Adoption of AI-supported suggestions. &
\noindent Accepted AI suggestions relative to all AI suggestions. \tabularnewline

Change Validation Rate &
Tests, CI/CD &
\noindent Changes verified before production. &
\noindent Share of changes with successful tests and quality gates. \tabularnewline

Human Override Rate &
Agent and review logs &
\noindent Human interventions in AI workflows. &
\noindent Corrections, cancellations, or overrides per agentic task. \tabularnewline

Autonomous Task Completion &
Agent logs &
\noindent Autonomously completed tasks. &
\noindent Agentic tasks without intervention up to successful merge. \tabularnewline

Rollback Rate &
Deployment and release systems &
\noindent Reverted changes. &
\noindent Production deployments followed by rollback. \tabularnewline

Cost per Task &
API and usage metrics &
\noindent Cost per development task. &
\noindent Token, API, and runtime cost per ticket, task, or PR. \tabularnewline

Audit Coverage &
Audit and logging systems &
\noindent Traceability of AI-supported changes. &
\noindent AI changes with audit trail, model, user, and approval. \tabularnewline

\bottomrule
\end{tabularx}
\end{table}

\FloatBarrier

\FloatBarrier

\clearpage
\phantomsection

\section*{B\quad Illustrative Claude-Oriented Codebase Harness}
\addcontentsline{toc}{section}{B Illustrative Claude-Oriented Codebase Harness}
\label{app:codebase_harness}
\label{app:codebase_harness_structure}

\providecommand{\cbhmd}[1]{%
{\setlength{\fboxsep}{1pt}\colorbox{gray!14}{\strut\texttt{#1}}}%
}

This appendix complements the supplementary tables with an illustrative
example of how an AI harness can become visible in a concrete
repository structure. The example is not intended as a general
standard. It focuses on a Claude Code-oriented setup and therefore
places \texttt{CLAUDE.md} at the center as the main project-specific
instruction file. The general idea can be transferred to other AI
development tools, but concrete files and conventions remain
tool-dependent.

The appendix refers to a Claude Code-oriented setup. Claude Code
supports project-level memory through \texttt{CLAUDE.md}. Such a file
can contain shared project instructions, such as architecture notes,
coding standards, or recurring workflows
~\cite{anthropicclaudecode_memory}. In GitHub-based development
environments, workflow files under \texttt{.github/workflows/} are
typically defined as YAML files and can implement build, test,
security, and release gates~\cite{githubactions_workflows}.
Architecture Decision Records provide a lightweight way to document
architecture decisions and their rationale within the repository
~\cite{adrgithub}.

The repository itself can become part of the harness if it provides
relevant context in a structured and retrievable form. This includes
project-specific instructions, architecture and domain documentation,
security and operations knowledge, source code, tests, quality
scripts, and CI/CD workflows.
Figure~\ref{fig:codebase_harness_structure} shows such a structure in
compact form. The highlighted entries denote Markdown files that can
explicitly contribute to the harness by making project knowledge, tool
rules, security requirements, or operational expectations available.

\begin{figure}[H]
\centering
\begin{minipage}{0.92\textwidth}
\scriptsize
\ttfamily

my-software-system/\par
\hspace*{1em}|-- \cbhmd{README.md}\par
\hspace*{1em}|-- \cbhmd{CLAUDE.md}\par
\hspace*{1em}|-- .github/\par
\hspace*{2em}|~~`-- workflows/\par
\hspace*{3em}|~~~~|-- ci.yml\par
\hspace*{3em}|~~~~|-- security-scan.yml\par
\hspace*{3em}|~~~~`-- release.yml\par
\hspace*{1em}|-- docs/\par
\hspace*{2em}|~~|-- architecture/\par
\hspace*{3em}|~~~~|-- \cbhmd{overview.md}\par
\hspace*{3em}|~~~~`-- \cbhmd{adr/*.md}\par
\hspace*{2em}|~~|-- domain/\par
\hspace*{3em}|~~~~|-- \cbhmd{glossary.md}\par
\hspace*{3em}|~~~~`-- \cbhmd{business-rules.md}\par
\hspace*{2em}|~~|-- security/\par
\hspace*{3em}|~~~~|-- \cbhmd{data-classification.md}\par
\hspace*{3em}|~~~~`-- \cbhmd{ai-guardrails.md}\par
\hspace*{2em}|~~`-- operations/\par
\hspace*{3em}|~~~~|-- \cbhmd{runbook.md}\par
\hspace*{3em}|~~~~`-- \cbhmd{incident-handling.md}\par
\hspace*{1em}|-- src/\par
\hspace*{2em}|~~`-- \textit{application source tree}\par
\hspace*{1em}|-- tests/\par
\hspace*{2em}|~~`-- \textit{unit, integration, regression, and security tests}\par
\hspace*{1em}`-- scripts/\par
\hspace*{2em}~~~`-- \textit{quality, security, and AI-review scripts}

\end{minipage}
\caption{Illustrative Claude-oriented repository structure for making an AI harness visible in a codebase. Highlighted entries denote Markdown files that explicitly support AI context, rules, domain knowledge, security requirements, operations knowledge, or review preparation.}
\label{fig:codebase_harness_structure}
\end{figure}

The files shown do not constitute the entire harness, but only the
part of it that is visible in the repository. A complete harness also
includes the execution environment, access to tools, the selection and
provisioning of context, permission boundaries, quality gates,
logging, and human approvals. The repository structure therefore
mainly ensures that relevant rules, domain information, and
verification requirements become findable and reusable for AI tools.

Table~\ref{tab:codebase_markdown_files} describes the Markdown files
highlighted in Figure~\ref{fig:codebase_harness_structure}. The
descriptions are exemplary and should be adapted to the specific tool
environment, architecture, domain, and governance requirements of a
project.

\begin{table}[H]
\caption{Exemplary purpose of Markdown files in the illustrative codebase harness}
\label{tab:codebase_markdown_files}
\centering
\scriptsize
\renewcommand{\arraystretch}{1.18}
\setlength{\tabcolsep}{4pt}
\rowcolors{2}{gray!8}{white}

\begin{tabularx}{\textwidth}{p{5.5cm} X}
\toprule
\textbf{Markdown file} &
\textbf{Exemplary purpose in the harness} \tabularnewline
\midrule

\texttt{README.md} &
Provides the general entry point to the project, including system purpose, setup information, basic development commands, and references to relevant documentation. \tabularnewline

\texttt{CLAUDE.md} &
Defines project-specific instructions for Claude Code, such as architecture requirements, coding expectations, testing rules, security boundaries, and pull-request expectations. \tabularnewline

\texttt{docs/architecture/overview.md} &
Summarizes the system architecture, central components, dependencies, deployment context, and design requirements that are relevant for AI-supported changes. \tabularnewline

\texttt{docs/architecture/adr/*.md} &
Documents architecture decisions and their rationale so that AI tools and reviewers can understand why certain design decisions or boundaries exist. \tabularnewline

\texttt{docs/domain/glossary.md} &
Defines domain terms and supports consistent naming, requirements interpretation, documentation, and generated code in line with the domain language. \tabularnewline

\texttt{docs/domain/business-rules.md} &
Documents central business rules and constraints that AI-generated changes should consider during analysis, implementation, and testing. \tabularnewline

\texttt{docs/security/data-classification.md} &
Describes data categories, sensitivity levels, and handling rules for privacy- and security-relevant development tasks. \tabularnewline

\texttt{docs/security/ai-guardrails.md} &
Defines project-specific guardrails for AI use, including restricted data, prohibited actions, required reviews, and rules for production-proximate changes. \tabularnewline

\texttt{docs/operations/runbook.md} &
Describes operational procedures, typical failure patterns, recovery steps, and escalation paths that can support AI-assisted incident analysis and operations support. \tabularnewline

\texttt{docs/operations/incident-handling.md} &
Describes how incidents are classified, analyzed, documented, and resolved, thereby supporting AI-assisted triage and post-incident learning. \tabularnewline

\bottomrule
\end{tabularx}
\end{table}

In this example, \texttt{CLAUDE.md} serves as the primary
tool-specific instruction file. Claude Code can use such
project-specific instructions as persistent guidance for AI-supported
development work~\cite{anthropicclaudecode_memory}. The file can
describe architecture boundaries, coding conventions, testing
expectations, security requirements, and review rules.
Figure~\ref{fig:claude_md_example} shows an illustrative example. It
is intentionally generic. It reflects typical categories of
project-specific instructions, but it is not a tool-independent
standard.

\providecommand{\claudeexrowgray}[1]{%
\noindent\colorbox{gray!8}{%
\parbox{\dimexpr\linewidth-2\fboxsep\relax}{#1}}%
\par\vspace{0.35em}
}

\providecommand{\claudeexrowwhite}[1]{%
\noindent\parbox{\linewidth}{#1}%
\par\vspace{0.35em}
}

\begin{figure}[H]
\centering
\small
\begin{minipage}{0.88\textwidth}

\claudeexrowwhite{%
\texttt{Project Instructions}
}

\claudeexrowgray{%
{\scriptsize
\textbf{Architecture context.}
Use \texttt{docs/architecture/} and existing Architecture Decision
Records before proposing structural changes.
}
}

\claudeexrowwhite{%
{\scriptsize
\textbf{System boundaries.}
Respect the separation between the central application areas. Do not
move responsibilities across these areas without an explicit rationale.
}
}

\claudeexrowgray{%
{\scriptsize
\textbf{Coding conventions.}
Follow existing coding conventions, naming patterns, and domain terms.
Prefer consistency with nearby code over introducing new style variants.
}
}

\claudeexrowwhite{%
{\scriptsize
\textbf{Tests.}
Add or update tests for functional changes. Use existing unit,
integration, or regression test structures where possible.
}
}

\claudeexrowgray{%
{\scriptsize
\textbf{Quality checks.}
Run available quality and security checks before proposing a pull
request. Report which checks were executed and which checks remain
open.
}
}

\claudeexrowwhite{%
{\scriptsize
\textbf{Security relevant changes.}
Do not change authentication, authorization, encryption, dependency
handling, deployment settings, or permissions without explicit human
review.
}
}

\claudeexrowgray{%
{\scriptsize
\textbf{Change size.}
Prefer small and reversible changes on separate branches. Do not mix
refactoring, feature work, and security relevant changes in one
proposal.
}
}

\claudeexrowwhite{%
{\scriptsize
\textbf{Pull request summary.}
Before proposing a pull request, summarize the affected files,
executed tests, remaining risks, and assumptions.
}
}

\end{minipage}
\caption{Illustrative example of project-specific AI instructions in \texttt{CLAUDE.md}.}
\label{fig:claude_md_example}
\end{figure}

The surrounding repository structure provides additional context and
control. Architecture documentation and Architecture Decision Records
support design consistency. Domain documentation helps align generated
changes with domain language and business rules. Test directories,
quality scripts, security scans, and CI/CD workflows define the
verification path for AI-supported changes. Operations documentation
connects development work with deployment, incident handling, and
production-related constraints.

The \texttt{.yml} files in the workflow directory are not Claude
instruction files. They represent CI/CD and automation mechanisms
through which build steps, test execution, security scans, release
procedures, and further quality gates can be defined. In the harness,
they therefore contribute to verification and control, not to prompt
or instruction management.

Such a structure does not make the codebase autonomous. It creates a
controlled working environment for AI-supported and agentic
development. AI tools can retrieve relevant context, propose changes
on branches, execute tests, and prepare pull requests. The harness
keeps these activities connected to quality gates, permission
boundaries, audit logs, and human review.

The example therefore illustrates only one possible form of a
codebase harness. In practice, it must be complemented by operational
mechanisms. These include sandboxes, branch rules, CI/CD gates, tool
permissions, audit logs, cost control, and defined approval steps.
Only the combination of these elements turns a documented repository
structure into a controllable harness for AI-assisted and agentic
development work.


\begin{thebibliography}{99}
\bibitem{hou2024llmse} X.~Hou, Y.~Wang, et al., ``Large Language Models for Software Engineering: A Systematic Literature Review,'' \emph{ACM Computing Surveys}, 2024.
\bibitem{fan2023survey} Z.~Fan et al., ``A Survey on Large Language Models for Software Engineering,'' \emph{arXiv preprint arXiv:2308.10620}, 2023.
\bibitem{terragni2025future} V.~Terragni, A.~Vella, P.~Roop, and K.~Blincoe, ``The Future of AI-driven Software Engineering,'' \emph{ACM Transactions on Software Engineering and Methodology}, 2025.
\bibitem{chen2021codex} M.~Chen, J.~Tworek, H.~Jun, et al., ``Evaluating Large Language Models Trained on Code,'' \emph{arXiv preprint arXiv:2107.03374}, 2021.
\bibitem{roziere2023codellama} B.~Rozi{\`e}re et al., ``Code Llama: Open Foundation Models for Code,'' \emph{arXiv preprint arXiv:2308.12950}, 2023.
\bibitem{li2023starcoder} R.~Li, L.~Ben Allal, et al., ``StarCoder: May the Source Be With You!,'' \emph{arXiv preprint arXiv:2305.06161}, 2023.
\bibitem{hamza2023humanai} A.~Hamza et al., ``Human-AI Collaboration in Software Engineering,'' in \emph{Proc. Int. Workshop on Human-Centered AI for Software Engineering}, 2023.
\bibitem{treude2025taxonomy} C.~Treude et al., ``A Taxonomy of Human-AI Interaction in Software Engineering,'' \emph{Empirical Software Engineering}, 2025.
\bibitem{barke2023groundedcopilot} S.~Barke, M.~B. James, and N.~Polikarpova, ``Grounded Copilot: How Programmers Interact with Code-Generating Models,'' in \emph{Proc. ACM Program. Lang. (OOPSLA)}, 2023.
\bibitem{vaithilingam2022expectation} P.~Vaithilingam et al., ``Expectation vs. Experience: Evaluating the Usability of Code Generation Tools Powered by Large Language Models,'' in \emph{CHI Extended Abstracts}, 2022.
\bibitem{mozannar2024reading} H.~Mozannar et al., ``Reading Between the Lines: Modeling User Behavior and Costs in AI-assisted Programming,'' in \emph{Proc. CHI Conference on Human Factors in Computing Systems}, 2024.
\bibitem{russo2024adoption} D.~Russo, ``Navigating the Complexity of Generative AI Adoption in Software Engineering,'' \emph{ACM Transactions on Software Engineering and Methodology}, 2024.
\bibitem{khemka2024support}
M. Khemka and B. Houck,
``Toward Effective AI Support for Developers: A Survey of Desires and
Concerns,''
\emph{Communications of the ACM}, vol. 67, no. 11, pp. 42--49, 2024,
doi: 10.1145/3690928.
\bibitem{peng2023copilot} S.~Peng, E.~Kalliamvakou, et al., ``The Impact of AI on Developer Productivity: Evidence from GitHub Copilot,'' \emph{arXiv preprint arXiv:2302.06590}, 2023.
\bibitem{cui2025highskilled} Z.~Cui et al., ``Generative AI and Productivity in High-Skilled Work,'' \emph{Management Science}, 2025.
\bibitem{metr2025productivity} J.~Becker, N.~Rush, B.~Barnes, and D.~Rein, ``Measuring the Impact of Early-2025 AI on Experienced Open-Source Developer Productivity,'' \emph{arXiv preprint arXiv:2507.09089}, 2025.
\bibitem{liang2024usability} J.~T. Liang, C.~Yang, and B.~A. Myers, ``A Large-Scale Survey on the Usability of AI Programming Assistants: Successes and Challenges,'' in \emph{Proc. CHI Conference on Human Factors in Computing Systems}, 2024.
\bibitem{sergeyuk2024practice} A.~Sergeyuk, Y.~Golubev, T.~Bryksin, and I.~Ahmed, ``Using AI-Based Coding Assistants in Practice: State of Affairs, Perceptions, and Ways Forward,'' \emph{arXiv preprint arXiv:2406.07765}, 2024.
\bibitem{github2024codequality} GitHub, ``Does GitHub Copilot Improve Code Quality? Here's What the Data Says,'' 2024. Available: \url{https://github.blog/news-insights/research/does-github-copilot-improve-code-quality-heres-what-the-data-says/}, accessed 2026-05-20.
\bibitem{dora2024stateofdevops} Google Cloud DORA, ``2024 State of DevOps Report,'' 2024. Available: \url{https://dora.dev/research/2024/dora-report/}, accessed 2026-05-20.
\bibitem{pearce2022asleep} H.~Pearce, B.~Ahmad, et al., ``Asleep at the Keyboard? Assessing the Security of GitHub Copilot's Code Contributions,'' in \emph{IEEE Symposium on Security and Privacy}, 2022.
\bibitem{sandoval2023lostatc} G.~Sandoval et al., ``Lost at C: A User Study on the Security Implications of LLM Code Assistants,'' in \emph{USENIX Security Symposium}, 2023.
\bibitem{khoury2023securechatgpt} R.~Khoury et al., ``How Secure is Code Generated by ChatGPT?,'' \emph{IEEE Access}, 2023.
\bibitem{liu2024evalplus} J.~Liu et al., ``Is Your Code Generated by ChatGPT Really Correct? Rigorous Evaluation of Large Language Models for Code Generation,'' \emph{Advances in Neural Information Processing Systems}, 2024.
\bibitem{ouyang2024nondeterminism} Y.~Ouyang et al., ``An Empirical Study of the Non-determinism of ChatGPT in Code Generation,'' \emph{arXiv preprint arXiv:2402.09176}, 2024.
\bibitem{spracklen2024packagehallucinations} M.~Spracklen et al., ``We Have a Package for You! A Comprehensive Analysis of Package Hallucinations by Code Generating LLMs,'' \emph{arXiv preprint arXiv:2406.10279}, 2024.
\bibitem{jimenez2024swebench} C.~Jimenez et al., ``SWE-bench: Can Language Models Resolve Real-World GitHub Issues?,'' in \emph{International Conference on Learning Representations}, 2024.
\bibitem{yang2024sweagent} J.~Yang et al., ``SWE-agent: Agent-Computer Interfaces Enable Automated Software Engineering,'' \emph{arXiv preprint arXiv:2405.15793}, 2024.
\bibitem{hong2024metagpt} S.~Hong et al., ``MetaGPT: Meta Programming for a Multi-Agent Collaborative Framework,'' in \emph{International Conference on Learning Representations}, 2024.
\bibitem{wu2024autogen} Q.~Wu et al., ``AutoGen: Enabling Next-Gen LLM Applications via Multi-Agent Conversation,'' \emph{Microsoft Research Technical Report}, 2024.
\bibitem{wang2024openhands} Q.~Wang et al., ``OpenHands: An Open Platform for AI Software Developers as Generalist Agents,'' \emph{arXiv preprint arXiv:2407.16741}, 2024.
\bibitem{zhang2024agents} C.~Zhang et al., ``A Survey on Large Language Model Based Autonomous Agents,'' \emph{Frontiers of Computer Science}, 2024.
\bibitem{zadenoori2025requirements} M.~Zadenoori et al., ``Large Language Models in Requirements Engineering: A Systematic Literature Review,'' \emph{Requirements Engineering}, 2025.
\bibitem{jiang2024codegeneration} N.~Jiang et al., ``A Survey on Code Generation with Large Language Models,'' \emph{ACM Computing Surveys}, 2024.
\bibitem{chen2024codeevaluation} Y.~Chen et al., ``Evaluation Methods for AI-Based Code Generation: A Survey,'' \emph{IEEE Access}, 2024.
\bibitem{xia2023apr} C.~Xia et al., ``Automated Program Repair with Large Language Models,'' \emph{IEEE Transactions on Software Engineering}, 2023.
\bibitem{zhang2025providerbias} X.~Zhang, J.~Zhai, S.~Ma, Q.~Bao, W.~Jiang, C.~Shen, and Y.~Liu, ``Unveiling Provider Bias in Large Language Models for Code Generation,'' in \emph{Proc. ACL}, 2025.
\bibitem{rumpe2010generative} B.~Rumpe, M.~Schindler, S.~V{\"o}lkel, and I.~Weisem{\"o}ller, ``Generative Software Development,'' in \emph{Proc. 32nd ACM/IEEE International Conference on Software Engineering, Volume 2}, pp.~473--474, 2010.
\bibitem{zhang2024agilegen} S.~Zhang, Z.~Xing, R.~Guo, F.~Xu, L.~Chen, Z.~Zhang, X.~Zhang, Z.~Feng, and Z.~Zhuang, ``Empowering Agile-Based Generative Software Development through Human-AI Teamwork,'' \emph{arXiv preprint arXiv:2407.15568}, 2024.
\bibitem{hassan2024ainative} A.~E. Hassan, G.~A. Oliva, D.~Lin, B.~Chen, and Z.~M. Jiang, ``Towards AI-Native Software Engineering (SE 3.0): A Vision and a Challenge Roadmap,'' \emph{arXiv preprint arXiv:2410.06107}, 2024.
\bibitem{li2025aiteammates} H.~Li, H.~Zhang, and A.~E. Hassan, ``The Rise of AI Teammates in Software Engineering (SE) 3.0: How Autonomous Coding Agents Are Reshaping Software Engineering,'' \emph{arXiv preprint arXiv:2507.15003}, 2025.
\bibitem{aws2025addlc} Amazon Web Services, ``AI-Driven Development Life Cycle: Reimagining Software Engineering,'' AWS DevOps \& Developer Productivity Blog, 2025. Available: \url{https://aws.amazon.com/blogs/devops/ai-driven-development-life-cycle/}, accessed 2026-05-24.
\bibitem{githubcopilotdocs} GitHub, ``GitHub Copilot Documentation,'' 2025. Available: \url{https://docs.github.com/copilot}, accessed 2026-05-20.
\bibitem{githubcloudagent} GitHub, ``About GitHub Copilot Coding Agent,'' 2025. Available: \url{https://docs.github.com/copilot/concepts/agents/coding-agent/about-coding-agent}, accessed 2026-05-20.
\bibitem{cursoragent} Cursor, ``Cursor Agent,'' 2025. Available: \url{https://docs.cursor.com/agent}, accessed 2026-05-20.
\bibitem{anthropicclaudecode} Anthropic, ``Claude Code,'' 2025. Available: \url{https://docs.anthropic.com/en/docs/claude-code/overview}, accessed 2026-05-20.
\bibitem{amazon2024qdeveloper} Amazon Web Services, ``Amazon Q Developer,'' 2024. Available: \url{https://aws.amazon.com/q/developer/}, accessed 2026-05-20.
\bibitem{github2024accenture} GitHub and Accenture, ``Research: Quantifying GitHub Copilot's Impact in the Enterprise,'' 2024. Available: \url{https://github.blog/news-insights/research/research-quantifying-github-copilots-impact-in-the-enterprise-with-accenture/}, accessed 2026-05-20.
\bibitem{google2024q3} Alphabet, ``Q3 2024 Earnings Call,'' 2024. Available: \url{https://abc.xyz/investor/}, accessed 2026-05-20.
\bibitem{shopify2025memo} Shopify, ``Internal Memo on AI Expectations,'' 2025. Available: \url{https://www.shopify.com/}, accessed 2026-05-20.
\bibitem{theverge2025shopify} The Verge, ``Shopify Memo Signals AI-first Expectations,'' 2025. Available: \url{https://www.theverge.com/}, accessed 2026-05-20.
\bibitem{openai2025gpt41} OpenAI, ``Introducing GPT-4.1 in the API,'' 2025. Available: \url{https://openai.com/index/gpt-4-1/}, accessed 2026-05-24.
\bibitem{openai2025gpt5} OpenAI, ``Introducing GPT-5,'' 2025. Available: \url{https://openai.com/index/introducing-gpt-5/}, accessed 2026-05-24.
\bibitem{anthropic2025claude4} Anthropic, ``Introducing Claude 4,'' 2025. Available: \url{https://www.anthropic.com/news/claude-4}, accessed 2026-05-24.
\bibitem{google2025gemini25pro} Google DeepMind, ``Gemini 2.5: Our most intelligent AI model,'' 2025. Available: \url{https://blog.google/innovation-and-ai/models-and-research/google-deepmind/gemini-model-thinking-updates-march-2025/}, accessed 2026-05-24.
\bibitem{mistral2024codestral} Mistral AI, ``Codestral,'' 2024. Available: \url{https://mistral.ai/news/codestral}, accessed 2026-05-24.
\bibitem{guo2024deepseekcoder} D.~Guo et al., ``DeepSeek-Coder: When the Large Language Model Meets Programming -- The Rise of Code Intelligence,'' \emph{arXiv preprint arXiv:2401.14196}, 2024.
\bibitem{hui2024qwen25coder} B.~Hui et al., ``Qwen2.5-Coder Technical Report,'' \emph{arXiv preprint arXiv:2409.12186}, 2024.
\bibitem{dora2025aicapabilities} Google Cloud DORA, ``Introducing the DORA AI Capabilities Model,'' 2025. Available: \url{https://cloud.google.com/blog/products/ai-machine-learning/introducing-doras-inaugural-ai-capabilities-model}, accessed 2026-05-24.
\bibitem{forrester2024turingbots} D.~Lo Giudice, ``The Future Is Now: TuringBots Will Collapse The Software Development Lifecycle Silos,'' Forrester Blog, 2024. Available: \url{https://www.forrester.com/blogs/the-future-is-now-turingbots-will-collapse-the-software-development-life-cycle-siloes/}, accessed 2026-05-24.
\bibitem{gartner2025softwaretrends} Gartner, ``Gartner Identifies the Top Strategic Trends in Software Engineering for 2025 and Beyond,'' 2025. Available: \url{https://www.gartner.com/en/newsroom/press-releases/2025-07-01-gartner-identifies-the-top-strategic-trends-in-software-engineering-for-2025-and-beyond}, accessed 2026-05-24.
\bibitem{mckinsey2025softwarevalue} McKinsey \& Company, ``Unlocking the value of AI in software development,'' 2025. Available: \url{https://www.mckinsey.com/industries/technology-media-and-telecommunications/our-insights/unlocking-the-value-of-ai-in-software-development}, accessed 2026-05-24.
\bibitem{thoughtworks2024aitools} B.~B{\"o}ckeler, ``Navigating the landscape of AI tools for software delivery,'' Thoughtworks, 2024. Available: \url{https://www.thoughtworks.com/insights/articles/ai-tools-software-delivery}, accessed 2026-05-24.
\bibitem{githubadvancedsecurity} GitHub, ``GitHub Advanced Security Documentation,'' 2025. Available: \url{https://docs.github.com/en/get-started/learning-about-github/about-github-advanced-security}, accessed 2026-05-24.
\bibitem{snyk2025docs} Snyk, ``Snyk Documentation,'' 2025. Available: \url{https://docs.snyk.io/}, accessed 2026-05-24.
\bibitem{semgrep2025docs} Semgrep, ``Semgrep Documentation,'' 2025. Available: \url{https://semgrep.dev/docs/}, accessed 2026-05-24.
\bibitem{trivy2025docs} Aqua Security, ``Trivy Documentation,'' 2025. Available: \url{https://trivy.dev/latest/docs/}, accessed 2026-05-24.
\bibitem{sonarqube2025aicodeassurance} SonarSource, ``AI Code Assurance,'' 2025. Available: \url{https://docs.sonarsource.com/sonarqube-server/latest/ai-code-assurance/}, accessed 2026-05-24.
\bibitem{hashimoto2026aiadoption} M.~Hashimoto, ``My AI Adoption Journey,'' 2026. Available: \url{https://mitchellh.com/writing/my-ai-adoption-journey}, accessed 2026-06-07.
\bibitem{openai2026harness} R.~Lopopolo, ``Harness Engineering: Leveraging Codex in an Agent-First World,'' OpenAI Engineering, 2026. Available: \url{https://openai.com/index/harness-engineering/}, accessed 2026-06-07.
\bibitem{bockeler2026harness} B.~B{\"o}ckeler, ``Harness Engineering for Coding Agent Users,'' Martin Fowler, 2026. Available: \url{https://martinfowler.com/articles/exploring-gen-ai/harness-engineering.html}, accessed 2026-06-07.
\bibitem{zhong2026aiharness} H.~Zhong and S.~Zhu, ``AI Harness Engineering: A Runtime Substrate for Foundation-Model Software Agents,'' \emph{arXiv preprint arXiv:2605.13357}, 2026.
\bibitem{norheim2024llmre} J.~J. Norheim, E.~Rebentisch, D.~Xiao, L.~Draeger, A.~Kerbrat, and O.~L. de Weck, ``Challenges in Applying Large Language Models to Requirements Engineering Tasks,'' \emph{Design Science}, vol.~10, e16, 2024, doi: 10.1017/dsj.2024.8.
\bibitem{mavin2009ears} A.~Mavin, P.~Wilkinson, A.~Harwood, and M.~Novak, ``Easy Approach to Requirements Syntax (EARS),'' in \emph{Proceedings of the 17th IEEE International Requirements Engineering Conference (RE)}, 2009, pp.~317--322, doi: 10.1109/RE.2009.9.
\bibitem{fischbach2022cira} J.~Fischbach et al., ``Automatic Creation of Acceptance Tests by Extracting Conditionals from Requirements: NLP Approach and Case Study,'' \emph{arXiv preprint arXiv:2202.00932}, 2022.
\bibitem{wang2019acceptancetests} C.~Wang, F.~Pastore, A.~Goknil, and L.~C. Briand, ``Automatic Generation of Acceptance Test Cases from Use Case Specifications: An NLP-Based Approach,'' \emph{arXiv preprint arXiv:1907.08490}, 2019.
\bibitem{sethupathy2023zerotouchdevops} U.~K.~A. Sethupathy, ``Zero-touch DevOps: A GenAI-orchestrated SDLC automation framework,'' \emph{World Journal of Advanced Engineering Technology and Sciences}, vol.~8, no.~2, pp.~420--433, 2023, doi: 10.30574/wjaets.2023.8.2.0119.
\bibitem{xmainframe2024} A.~T.~V. Dau et al., ``XMainframe: A Large Language Model for Mainframe Modernization,'' \emph{arXiv preprint arXiv:2408.04660}, 2024.
\bibitem{ibm2024coboljava} A.~Kumar et al., ``Automated Validation of COBOL to Java Transformation,'' in \emph{Proceedings of the 39th IEEE/ACM International Conference on Automated Software Engineering (ASE)}, 2024.
\bibitem{gandhi2024coboljava} S.~Gandhi et al., ``Translation of Low-Resource COBOL to Logically Correct and Readable Java leveraging High-Resource Java Refinement,'' in \emph{Proceedings of the ICSE 2024 Workshop on Large Language Models for Code (LLM4Code)}, 2024.
\bibitem{solovyeva2025plsqljava} L.~Solovyeva et al., ``Leveraging LLMs for Automated Translation of Legacy Code: A Case Study on PL/SQL to Java Transformation,'' \emph{arXiv preprint arXiv:2508.19663}, 2025.

\bibitem{anthropicclaudecode_memory}
Anthropic, ``Manage Claude's memory,'' 2026.
Available: \url{https://docs.anthropic.com/en/docs/claude-code/memory},
accessed 2026-06-12.

\bibitem{githubactions_workflows}
GitHub, ``Workflow syntax for GitHub Actions,'' 2026.
Available: \url{https://docs.github.com/en/actions/reference/workflows-and-actions/workflow-syntax},
accessed 2026-06-12.

\bibitem{adrgithub}
Architectural Decision Records, ``Architectural Decision Records,'' 2026.
Available: \url{https://adr.github.io/},
accessed 2026-06-12.

\bibitem{pan2026rho}
W. Pan, S. Liu, C.-Y. Lin, J. Zeng, X. Tang, X. Zhou, Y. Lu, and X. Jia,
``Retrospective Harness Optimization: Improving LLM Agents via
Self-Preference over Trajectory Rollouts,'' 2026.
Available: \url{https://arxiv.org/abs/2606.05922},
accessed 2026-06-13.

\end{thebibliography}
\end{document}